\documentclass[acmsmall]{acmart}

\usepackage{booktabs}
\usepackage{graphicx,tabularx,array}
\usepackage[font=bf]{caption}
\usepackage{pifont}
\usepackage{framed}
\usepackage{soul}
\usepackage{booktabs}  
\usepackage{csquotes}
\usepackage{multirow}
\usepackage{array}
\usepackage{url}
\usepackage{color}
\usepackage{colortbl}
\usepackage{wasysym}
\MakeOuterQuote{"}

\usepackage{hyperref}

\AtBeginDocument{%
  \providecommand\BibTeX{{%
    \normalfont B\kern-0.5em{\scshape i\kern-0.25em b}\kern-0.8em\TeX}}}

\setcopyright{acmcopyright}
\copyrightyear{2020}
\acmYear{2020}
\acmDOI{}

\acmJournal{CSUR}
\acmVolume{0}
\acmNumber{0}
\acmArticle{0}
\acmMonth{0}

\begin{document}

\title{Security of Cloud FPGAs: A Survey}

\author{Chenglu Jin}
\email{chenglu.jin@nyu.edu}
\orcid{0000-0001-6306-8019}
\affiliation{%
  \institution{New York University}
  \streetaddress{370 Jay Street}
  \city{Brooklyn}
  \state{New York}
  \postcode{11201}
}
\author{Vasudev Gohil}
\email{gohil.vasudev@tamu.edu}
\affiliation{%
  \institution{Texas A\&M University}
  \streetaddress{400 Bizzell Street}
  \city{College Station}
  \state{Texas}
  \postcode{77843}
}

\author{Ramesh Karri}
\email{rkarri@nyu.edu}
\affiliation{%
  \institution{New York University}
  \streetaddress{370 Jay Street}
  \city{Brooklyn}
  \state{New York}
  \postcode{11201}
}

\author{Jeyavijayan Rajendran}
\email{jv.rajendran@tamu.edu}
\affiliation{%
  \institution{Texas A\&M University}
  \streetaddress{400 Bizzell Street}
  \city{College Station}
  \state{Texas}
  \postcode{77843}
}

\renewcommand{\shortauthors}{Jin, et al.}

\begin{abstract}
Integrating Field Programmable Gate Arrays (FPGAs) with cloud computing instances is a rapidly emerging trend on commercial cloud computing platforms such as Amazon Web Services (AWS), Huawei cloud, and Alibaba cloud. Cloud FPGAs allow cloud users to build hardware accelerators to speed up the computation in the cloud. However, since the cloud FPGA technology is still in its infancy, the security implications of this integration of FPGAs in the cloud are not clear. In this paper, we survey the emerging field of cloud FPGA security, providing a comprehensive overview of the security issues related to cloud FPGAs, and highlighting future challenges in this research area.
\end{abstract}



\maketitle

\section{Introduction}

The last few decades have witnessed tremendous growth in the need for high-speed computation in the clouds. Solely using CPUs and GPUs can no longer meet the increasing performance demand, in terms of latency, throughput, and efficiency. Due to this, FPGAs have been integrated into cloud computation platforms to allow users to customize their hardware accelerators (to accelerate computationally intensive tasks) in the clouds. Many commercial cloud providers have already integrated or are integrating FPGAs in their cloud services platforms, e.g., Amazon~\cite{amazon_f1}, Huawei~\cite{huawei}, Alibaba~\cite{Aliyun_spec}, Microsoft~\cite{Azure}, and Texas Advanced Computer Center~\cite{tacc}. Intel predicted in 2016 that one-third of the cloud computing instances would have an FPGA by 2020~\cite{fpga_market}. Users can use these cloud FPGAs to accelerate computationally intensive workloads like artificial intelligence tasks, software-defined networking, big data analytics, genomics, electronic design automation, and image and video processing~\cite{catapult,amazon_f1}.

Comparing with traditional CPU-based or GPU-based cloud computation, FPGAs offer unique advantages. In particular, FPGAs are an ideal platform to perform \textit{parallel} computation with \textit{flexible} datapath and control. Thus, they can speed up computation with high efficiency. We explain these features in detail below.
\begin{itemize}
\item FPGAs can support \textit{massive parallelism} in computation. For example, each FPGA on Amazon cloud (Xilinx UltraScale+ VU9P) has more than two million customer-accessible FPGA programmable logic cells~\cite{aws_slides}, and they can all run in parallel to accelerate computation. For example, the instances on Amazon cloud accelerate the computing time by up to 100$\times$~\cite{amazon_f1}.
\item FPGAs are highly \textit{flexible} in building a datapath with arbitrary width; e.g., if an application needs a 9-bit integer, the user can configure the datapath to exact 9 bits without underutilizing any computational resources, while in a CPU, it would require two bytes to store 9 bits. Moreover, it is easier to use FPGAs to build a customized state machine to control the computation on FPGAs, which is more efficient than using software for fine-grained controls.
\end{itemize}

In addition to these advantages, the cost of general-purpose commercial FPGA-based cloud computing instances is meager: one Amazon \textit{EC2 f1.2xlarge} instance, which has one FPGA board, can cost as low as \$0.76 per hour~\cite{amazon_f1}, and one basic Huawei FP1 instance costs about \$0.98 per hour~\cite{huawei}. Table~\ref{tab:business} presents a summary of the platforms provided by leading FPGA cloud providers. Users can choose proper specifications (e.g., the number and vendor of their FPGAs, the number of CPU cores, and the size of memory) for their cloud computation.

\begin{table*}[t]\centering
\caption{Comparison of cloud FPGA providers. The specifications and prices are based on~\cite{amazon_f1,huawei,Aliyun_price,Aliyun_spec,Azure,Nimbix} as of March 2020, and the prices have been converted into US dollars for easy comparison. *Microsoft Azure provides cloud FPGAs as hardware accelerators only for machine learning.} \label{tab:business}
\resizebox{\textwidth}{!}{
\begin{tabular}{|l|c|c|c|c|c|}
\hline
\textbf{Provider} & \textbf{\# FPGAs/instance}   & \textbf{\# virtual CPUs/instance}    & \textbf{Memory (GB)}   & \textbf{SSD (GB)}       & \textbf{Price/hour}    \\ \hline
Amazon   & 1/2/8 (Xilinx VU9P) & 8/16/64 & 122/244/976 & 470/940/3760 & \$0.76 + \\ \hline
Huawei   & 1/2/4/8 (Xilinx VU9P) & 8/32/64 & 88/224/352/448/708  & N/A            & \$0.98 + \\ \hline
Alibaba  & 1/2/4 (Intel Arria 10 GX 1150 or Xilinx VU9P) & 4/8/16/28/32/56/64 & 16/32/60/64/112/120/128/224/256  & N/A            & \$0.14 + \\ \hline
Nimbix & 1 (Xilinx Alevo U50/U200/U250/U280) & 16/32/64 & 128 & N/A & \$3.00 + \\ \hline
Azure ML$^*$ & 1/2/4 (Intel Arria 10) & N/A & 112/224/448 & N/A & \$0.33 + \\ \hline
\end{tabular}}
\end{table*}

\begin{table*}[b!]\centering
\caption{Comparison between cloud security, FPGA security, and cloud FPGA security.}\label{tab:comp}
\resizebox{\textwidth}{!}{
\begin{tabular}{|l|c|c|c|c|}
\hline
                    & Assets Under Attacks & Threat Models &  Physical Accesses & Programmable Hardware \\ \hline
Cloud Security  & Data & Cloud \& Clients  & N                 & N                                              \\ \hline
FPGA Security  & Data \& H/W Design & H/W Users \& H/W Design Supply Chain    & Y                 & Y                                              \\ \hline
Cloud FPGA Security & Data \& H/W Design & Cloud \& Clients \& H/W Design Supply Chain & N                 & Y           \\ \hline
\end{tabular}}
\end{table*}

FPGA-accelerated clouds can be beneficial to a large variety of sectors. Deep learning technology has a wide range of applications. FPGA accelerators can boost the performance of deep learning technology, and thus accelerate numerous applications and services ranging from database management to artificial intelligence~\cite{zhang2015optimizing,guo2017survey}. Microsoft Brainwave project has developed a deep neural network architecture that can be synthesized on FPGAs to achieve ten to over thirty-five teraflops~\cite{brainwave,fowers2018configurable}. Also, FPGA accelerators can help speed up heavy computation tasks on video classification and genome analysis, as these algorithms have a tremendous amount of parallelism that can be exploited~\cite{chen2016spark,suda2016throughput,wei2019fpga}. FPGA accelerators that provide over ten times speedup in genome sequencing analysis are deployed on Amazon AWS F1 instances by Edico Genome~\cite{edico_genome}.  FPGAs can also enhance database management systems.  In particular, set-oriented queries in database systems are suitable for FPGA computation, as a high degree of parallelism exists in set data queries~\cite{mueller2009fpga}. More complex analytic operations of data have been accelerated by FPGA platforms tremendously~\cite{sukhwani2012database}. \textit{Bing} is currently powered by FPGA accelerators, which offer a 50\% improvement in throughput and a 25\% reduction in latency~\cite{catapult}. 

Building programmable hardware in the clouds improves the performance of cloud-hosted services significantly. However, this integration opens a new attack surface from an attacker's perspective. This is because FPGAs allow users to implement custom logic on them, unlike CPUs, and GPUs. A variety of attacks have been demonstrated in recent research papers~\cite{schellenberg2018inside,zhao2018fpga}, and researchers are developing countermeasures to thwart the attacks on cloud FPGAs~\cite{krautter2019mitigating,provelengios2019characterizing}. This paper surveys the broad landscape of cloud FPGA security research. It summarizes the state-of-the-art research and points out future research directions.%

We organize the whole paper in a way that it answers four fundamental research questions in cloud FPGA security research one by one:

\begin{enumerate}
    \item What are the security threats when FPGAs are introduced in a cloud platform?
    \item In what different ways can a malicious user attack an FPGA in the cloud?
    \item How to defend against such attacks on cloud FPGAs?
    \item How can we use FPGAs as a tool to enhance cloud security? 
\end{enumerate}

To better understand the differences between cloud security, FPGA security, and cloud FPGA security, we create Table~\ref{tab:comp} to show the comparison. Most importantly, the threat models of these three security research areas are different, and cloud FPGAs have the largest attack surface. In general, in the scenarios of cloud computing (cloud security and cloud FPGA security), we do not assume that users (either attackers or victims) have physical access to the computation resources. Additionally, traditional cloud security research does not assume that the underlying hardware can be maliciously altered by attackers (except the case of hardware Trojans). But with programmable hardware in the clouds, an attacker can create a hardware foothold in the system to launch attacks that were not possible before, e.g., side-channel attacks. In terms of the assets that defenders need to protect, hardware designs on FPGAs are valuable targets for FPGA security and cloud FPGA security attackers, in addition to the data that is computed or stored on the devices.

\vspace{2mm}
\noindent\textbf{Organization.} We introduce the background knowledge and the threat models of cloud FPGAs in Section~\ref{sec:background} and Section~\ref{sec:threat}, respectively. We survey the literature on attacks for a variety of threat models in Section~\ref{sec:taxonomy}. As there is a vast amount of research on power-based side-channel attacks and ring oscillator (RO) design variants, we provide the two case studies in Section~\ref{sec:case}. We discuss the countermeasures against the above attacks in Section~\ref{sec:counter}. Researchers have introduced various methods to use FPGAs to enhance system security (i.e., the security of cloud computation), which is presented in Section~\ref{sec:support}. The recent related surveys and the differences between our paper and other surveys are discussed in Section~\ref{sec:related}. We share our thoughts on future challenges and provide concluding remarks in Section~\ref{sec:future}. We categorize existing research on attacking or protecting cloud FPGAs in Table~\ref{tab:summary} as a systematic review.
\begin{table}[ht]
\centering
\caption{Categorization of cloud FPGA literature based on (1) threat model, (2) attack class, and (3) whether the study is about attack or defense or both.}\label{tab:summary}
\resizebox{0.873\textwidth}{!}{%
\begin{tabular}{|lcccccccccccccc|}
\specialrule{1pt}{1pt}{1pt}
 & \multicolumn{2}{c|}{\textbf{}} & \multicolumn{4}{c|}{\textbf{Threat model}} & \multicolumn{8}{c|}{\textbf{Attack class}} \\
\textbf{Papers} & \rotatebox{90}{Attacks} & \multicolumn{1}{c|}{\rotatebox{90}{Defenses}} & {\rotatebox{90}{Clouds}} & \rotatebox{90}{Co-tenants} & \rotatebox{90}{IP providers} & \multicolumn{1}{c|}{\rotatebox{90}{FPGA tools}} & {\rotatebox{90}{Direct data leakage   }} & {\rotatebox{90}{IP theft}} & {\rotatebox{90}{Logic tampering}} & \rotatebox{90}{Side-channel} & \rotatebox{90}{Fault-injection} & \rotatebox{90}{Denial-of-service attacks} & \rotatebox{90}{RowHammer} & \rotatebox{90}{Covert channel}\\ \specialrule{0.8pt}{1pt}{1pt}
\rowcolor[HTML]{DFDFDF} 
Huffmire \textit{et al.}~\cite{huffmire2007moats} & & \multicolumn{1}{c|}{\checkmark} &  & \checkmark & \checkmark & \multicolumn{1}{c|}{} & \checkmark & & & \checkmark & & & & \multicolumn{1}{c|}{\checkmark}\\
Note \textit{et al.}~\cite{note2008bitstream} & \checkmark & \multicolumn{1}{c|}{} & \checkmark & &  & \multicolumn{1}{c|}{} & & \checkmark & & & & & & \multicolumn{1}{c|}{}\\
\rowcolor[HTML]{DFDFDF} 
Endo \textit{et al.}~\cite{endo2012efficient} & & \multicolumn{1}{c|}{\checkmark} & \checkmark & \checkmark & \checkmark & \multicolumn{1}{c|}{} & & & & & \checkmark & & & \multicolumn{1}{c|}{}\\
Benz \textit{et al.}~\cite{benz2012bil} & \checkmark & \multicolumn{1}{c|}{} & \checkmark & &  & \multicolumn{1}{c|}{} & & \checkmark & & & & & & \multicolumn{1}{c|}{}\\
\rowcolor[HTML]{DFDFDF} 
Gnad \textit{et al.}~\cite{gnad2017voltage} & \checkmark & \multicolumn{1}{c|}{} &  & \checkmark & & \multicolumn{1}{c|}{} & &  &  &  &  & \checkmark  & & \multicolumn{1}{c|}{} \\
Schellenberg \textit{et al.}~\cite{schellenberg2018inside} & \checkmark & \multicolumn{1}{c|}{} &  & \checkmark & & \multicolumn{1}{c|}{} & & & & \checkmark & & & & \multicolumn{1}{c|}{} \\
\rowcolor[HTML]{DFDFDF}
Gnad \textit{et al.}~\cite{gnad2018checking} & & \multicolumn{1}{c|}{\checkmark} & & \checkmark & \checkmark & \multicolumn{1}{c|}{} & & & & \checkmark & \checkmark & \checkmark & & \multicolumn{1}{c|}{\checkmark} \\
Schellenberg \textit{et al.}~\cite{schellenberg2018remote} & \checkmark & \multicolumn{1}{c|}{} & & \checkmark & & \multicolumn{1}{c|}{} & & & &  \checkmark & & & & \multicolumn{1}{c|}{}\\
\rowcolor[HTML]{DFDFDF}
Hategekimana \textit{et al.}~\cite{hategekimana2018secure} &  & \multicolumn{1}{c|}{\checkmark} & & \checkmark & & \multicolumn{1}{c|}{} & \checkmark & & & & & & & \multicolumn{1}{c|}{} \\
Yazdanshenas \textit{et al.}~\cite{yazdanshenas2018improving} &  & \multicolumn{1}{c|}{\checkmark} &  & \checkmark & & \multicolumn{1}{c|}{} & \checkmark & & & & & & &  \multicolumn{1}{c|}{}\\
\rowcolor[HTML]{DFDFDF} 
Zhao \textit{et al.}~\cite{zhao2018fpga} & \checkmark &  \multicolumn{1}{c|}{} & & \checkmark & & \multicolumn{1}{c|}{} & & & & \checkmark & & & &  \multicolumn{1}{c|}{}\\
Krautter \textit{et al.}~\cite{krautter2018fpgahammer} & \checkmark & \multicolumn{1}{c|}{} &  & \checkmark & & \multicolumn{1}{c|}{} & & & & & \checkmark & & & \multicolumn{1}{c|}{}\\
\rowcolor[HTML]{DFDFDF} 
Ramesh \textit{et al.}~\cite{ramesh2018fpga} & \checkmark & \multicolumn{1}{c|}{} &  & \checkmark & \checkmark & \multicolumn{1}{c|}{} &  & & & \checkmark & & & &  \multicolumn{1}{c|}{\checkmark} \\
Bag \textit{et al.}~\cite{bag2018cryptographically} &  &  \multicolumn{1}{c|}{\checkmark} & \checkmark &  &  & \multicolumn{1}{c|}{} & & \checkmark & & & &  & & \multicolumn{1}{c|}{}\\
\rowcolor[HTML]{DFDFDF} 
Provelengios \textit{et al.}~\cite{provelengios2019characterizing} & & \multicolumn{1}{c|}{\checkmark} &  & \checkmark &  & \multicolumn{1}{c|}{} & & & & & \checkmark & \checkmark & &  \multicolumn{1}{c|}{} \\
Tian \textit{et al.}~\cite{tian2019temporal} & \checkmark & \multicolumn{1}{c|}{}  & & & \checkmark & \multicolumn{1}{c|}{} & & &  &  & & &  & \multicolumn{1}{c|}{\checkmark}\\
\rowcolor[HTML]{DFDFDF} 
Sugawara \textit{et al.}~\cite{sugawara2019oscillator} & \checkmark & \multicolumn{1}{c|}{} & & \checkmark & \checkmark & \multicolumn{1}{c|}{} & & & & \checkmark & \checkmark & \checkmark &  &  \multicolumn{1}{c|}{\checkmark}\\
Alam \textit{et al.}~\cite{alam2019ram} & \checkmark & \multicolumn{1}{c|}{} & & \checkmark & & \multicolumn{1}{c|}{} & & & & & \checkmark & & & \multicolumn{1}{c|}{}\\
\rowcolor[HTML]{DFDFDF} 
Weissman \textit{et al.}~\cite{weissman2019jackhammer} & \checkmark & \multicolumn{1}{c|}{} & &  \checkmark & & \multicolumn{1}{c|}{} & & & & & &  & \checkmark & \multicolumn{1}{c|}{}\\
Krautter \textit{et al.}~\cite{krautter2019mitigating} &  & \multicolumn{1}{c|}{\checkmark} & & \checkmark & \checkmark & \multicolumn{1}{c|}{} & & & & \checkmark & \checkmark & \checkmark &  & 
\multicolumn{1}{c|}{\checkmark} \\
\rowcolor[HTML]{DFDFDF} 
Krautter \textit{et al.}~\cite{krautter2019active} &  & \multicolumn{1}{c|}{\checkmark} & \checkmark & \checkmark & \checkmark & \multicolumn{1}{c|}{} & & & & \checkmark & & & &  \multicolumn{1}{c|}{\checkmark}\\
Mahmoud \textit{et al.}~\cite{mahmoud2019timing} & \checkmark & \multicolumn{1}{c|}{} & & \checkmark & & \multicolumn{1}{c|}{} & & & & & \checkmark & & &   \multicolumn{1}{c|}{}\\
\rowcolor[HTML]{DFDFDF} 
Gnad \textit{et al.}~\cite{gnadvoltage} & \checkmark & \multicolumn{1}{c|}{} &  &  & \checkmark & \multicolumn{1}{c|}{} & & & & & & & & \multicolumn{1}{c|}{\checkmark}\\
Elnaggar \textit{et al.}~\cite{elnaggar2019multi} & \checkmark & \multicolumn{1}{c|}{\checkmark} & & \checkmark & & \multicolumn{1}{c|}{} & \checkmark & & \checkmark & & & & &  \multicolumn{1}{c|}{}\\
\rowcolor[HTML]{DFDFDF} 
Gravellier \textit{et al.}~\cite{gravellier2019remote} & \checkmark & \multicolumn{1}{c|}{} & & \checkmark & & \multicolumn{1}{c|}{} & & & & \checkmark & & & & \multicolumn{1}{c|}{}\\
Provelengios \textit{et al.}~\cite{provelengios2019characterization} & \checkmark & \multicolumn{1}{c|}{} & &  & \checkmark & \multicolumn{1}{c|}{} & & & & & & &  & \multicolumn{1}{c|}{\checkmark}\\
\rowcolor[HTML]{DFDFDF} 
Giechaskiel \textit{et al.}~\cite{giechaskielreading} & \checkmark & \multicolumn{1}{c|}{} &  &  & \checkmark & \multicolumn{1}{c|}{} & & &  &  & & & &  \multicolumn{1}{c|}{\checkmark}\\
Giechaskiel \textit{et al.}~\cite{giechaskiel2019leakier} & \checkmark & \multicolumn{1}{c|}{} &  &  & \checkmark & \multicolumn{1}{c|}{} & & &  &  & & & &  \multicolumn{1}{c|}{\checkmark}\\
\rowcolor[HTML]{DFDFDF} 
Giechaskiel \textit{et al.}~\cite{giechaskiel2019measuring} & \checkmark & \multicolumn{1}{c|}{} &  &  & \checkmark & \multicolumn{1}{c|}{} & & &  &  & & & &  \multicolumn{1}{c|}{\checkmark}\\
Luo \textit{et al.}~\cite{luo2019hill} &  &  \multicolumn{1}{c|}{\checkmark} &  &  & \checkmark & \multicolumn{1}{c|}{} & & & & & &  & & \multicolumn{1}{c|}{\checkmark}\\
\rowcolor[HTML]{DFDFDF} 
Matas \textit{et al.}~\cite{matasinvited} & \checkmark & \multicolumn{1}{c|}{\checkmark} & & \checkmark & & \multicolumn{1}{c|}{} & & & & \checkmark & \checkmark & \checkmark & & \multicolumn{1}{c|}{\checkmark}\\
Giechaskiel \textit{et al.}~\cite{giechaskiel2020capsule} & \checkmark & \multicolumn{1}{c|}{} & & & \checkmark & \multicolumn{1}{c|}{} & & & & & & & & \multicolumn{1}{c|}{\checkmark}\\
\rowcolor[HTML]{DFDFDF} 
Giechaskiel \textit{et al.}~\cite{giechaskiel2018leaky} & \checkmark & \multicolumn{1}{c|}{} & & & \checkmark & \multicolumn{1}{c|}{} & & & & & & & & \multicolumn{1}{c|}{\checkmark}\\
Glamocanin \textit{et al.}~\cite{glamocanin2020are} & \checkmark & \multicolumn{1}{c|}{} & & \checkmark & & \multicolumn{1}{c|}{} & & & & \checkmark & &  & & \multicolumn{1}{c|}{}\\
\rowcolor[HTML]{DFDFDF}
Krieg \textit{et al.} ~\cite{krieg2016malicious} & \checkmark & \multicolumn{1}{c|}{} & & & & \multicolumn{1}{c|}{\checkmark} &  & & \checkmark & & & & & \multicolumn{1}{c|}{\checkmark}\\
\specialrule{0.8pt}{1pt}{1pt}
\end{tabular}
}
\end{table}

\section{Background}\label{sec:background}

\begin{figure}
    \centering
    \includegraphics[width=0.6\textwidth, trim = 1.8cm 0 0 0, clip]{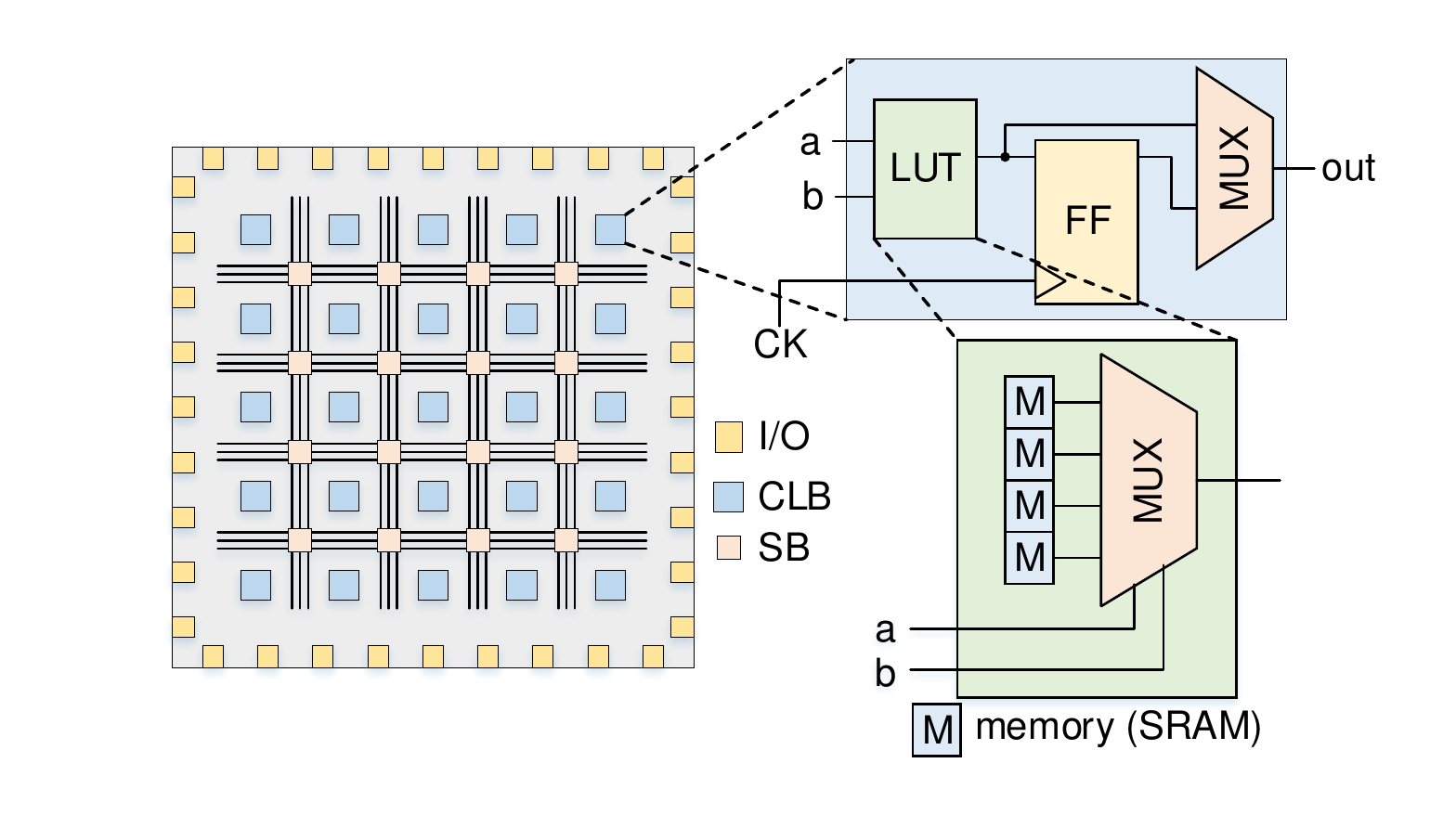}
    \caption{An FPGA has Configuration Logic Blocks (CLB), switch-boxes (SB), and input/output cells. Each CLB consists of a Look-Up Table (LUT), a flip-flop (FF), and a MUX.}
    \label{fig:FPGA_internals}
\end{figure}

\begin{figure*}[b!]
    \centering
    \includegraphics[width=\textwidth]{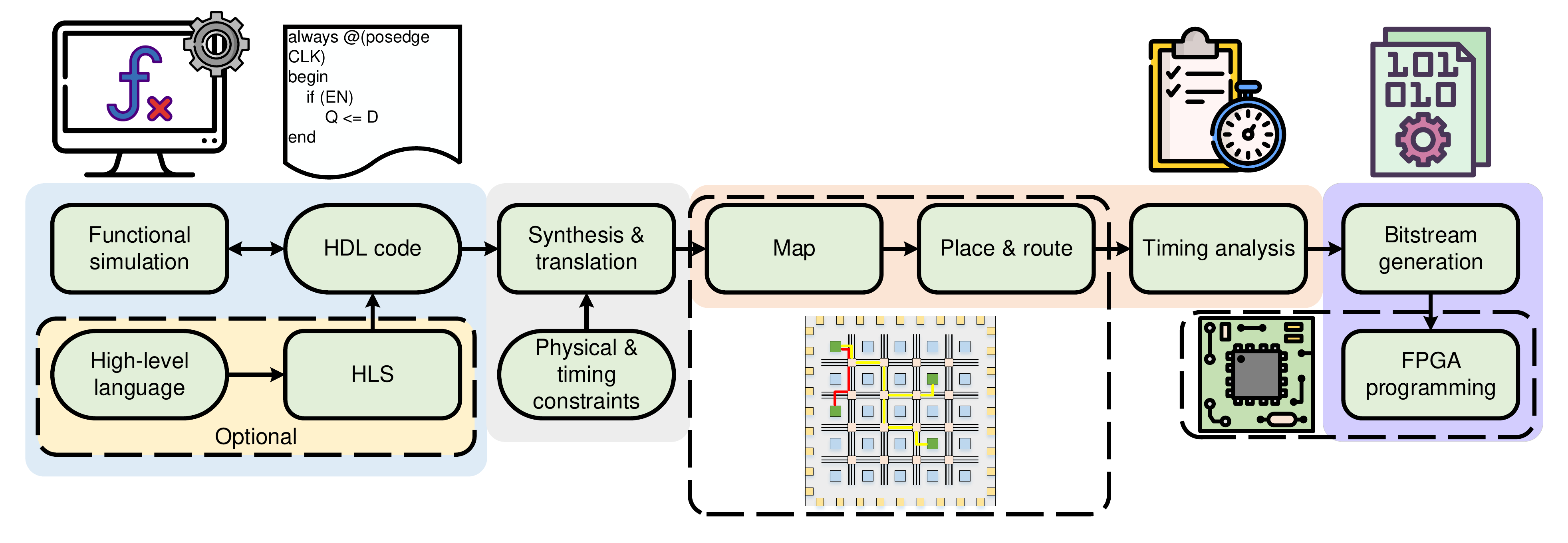}
    \caption{FPGA design flow: The yellow box indicates that the step inside that box is optional. The blue box shows the steps which give a HDL code as an output. The grey box indicates synthesis of the HDL code and its translation. The red box shows the mapping of the logic to LUTs, the placement and routing of LUTs, and the timing analysis step. The purple box indicates the steps to generate the final bitstream followed by programming the FPGA.}
    \label{fig:FPGA_Design_Flow}
\end{figure*}

\subsection{Field Programmable Gate Arrays}

FPGAs are integrated circuits composed of programmable blocks, allowing a user to program the circuit functionality as needed even after fabrication. Fig.~\ref{fig:FPGA_internals} shows a typical architecture of an FPGA. The architecture includes an array of configuration logic blocks (CLBs), switch boxes (SBs), and input/output pins. CLBs are composed of lookup tables (LUTs), flip-flops, and multiplexers. Each CLB can be programmed to implement any Boolean function with $n$ or fewer inputs, where $n$ is the input size of the LUT. The SBs in an FPGA can be configured to connect CLBs, so multiple CLBs can jointly construct a larger circuit and thus perform more complex computation. Input/output pins connect an FPGA with the outside world, such as power supply, clock signals, and other peripherals. 

\subsection{FPGA Design Flow}

Fig.~\ref{fig:FPGA_Design_Flow} shows the typical FPGA design flow. A designer designs the target system in terms of Hardware Description Language (HDL) codes, e.g., Verilog HDL or VHDL. After the HDL codes have been simulated and verified for correctness, they are synthesized and translated to a netlist by FPGA synthesis tools (e.g., Xilinx Vivado~\cite{vivado_hls} or Xilinx ISE~\cite{xilinx_ise}). The netlist describes how the hardware components, such as LUTs and registers, are connected.  The FPGA synthesis tool then maps the components to the actual hardware resources on a specified FPGA. Next, the routes between each component are optimized to meet the timing constraints and other physical constraints given by the user. The end goal of the design process is a bitstream file, which is a string of 0s and 1s. After the bitstream file is loaded onto an FPGA, the FPGA will function as intended by the user.

To reduce the design time and the verification efforts, a designer can specify the design using a high-level language (e.g., C or MATLAB). This also allows a developer, who does not have the required expertise in writing HDL codes, to use FPGAs for his/her need. This alternative is called High-Level Synthesis (HLS), and it can compile high-level programming language code to a functionally equivalent HDL code. Xilinx Vivado HLS~\cite{vivado_hls} and Intel High-level synthesis compiler~\cite{intel_hls} are examples of tools which provide this functionality. The FPGA synthesis tool processes the HDL code and creates a bitstream file used to program the FPGA.

The above two programming methods are available to the users of cloud FPGAs. So the users can either submit their hardware design as HDL codes or as a high-level language program. A user, even without much knowledge of hardware design, can start from a high-level language and run an HLS tool locally to create HDL codes for uploading to the clouds. The cloud service provider takes the source codes of user designs and integrates user logic with their IP cores (called shell on Amazon platforms) to build a bitstream file. This bitstream file is then loaded on an FPGA. In the current commercial setting, the cloud provider has full control over the compilation and deployment of user logic as it has to happen in an Amazon cloud node~\cite{aws_hdk}. 

\subsection{Architecture of FPGA clouds}

\begin{figure}[b!]
    \centering
    \includegraphics[width=\textwidth, trim = 0 0.5cm 0 0.5cm, clip]{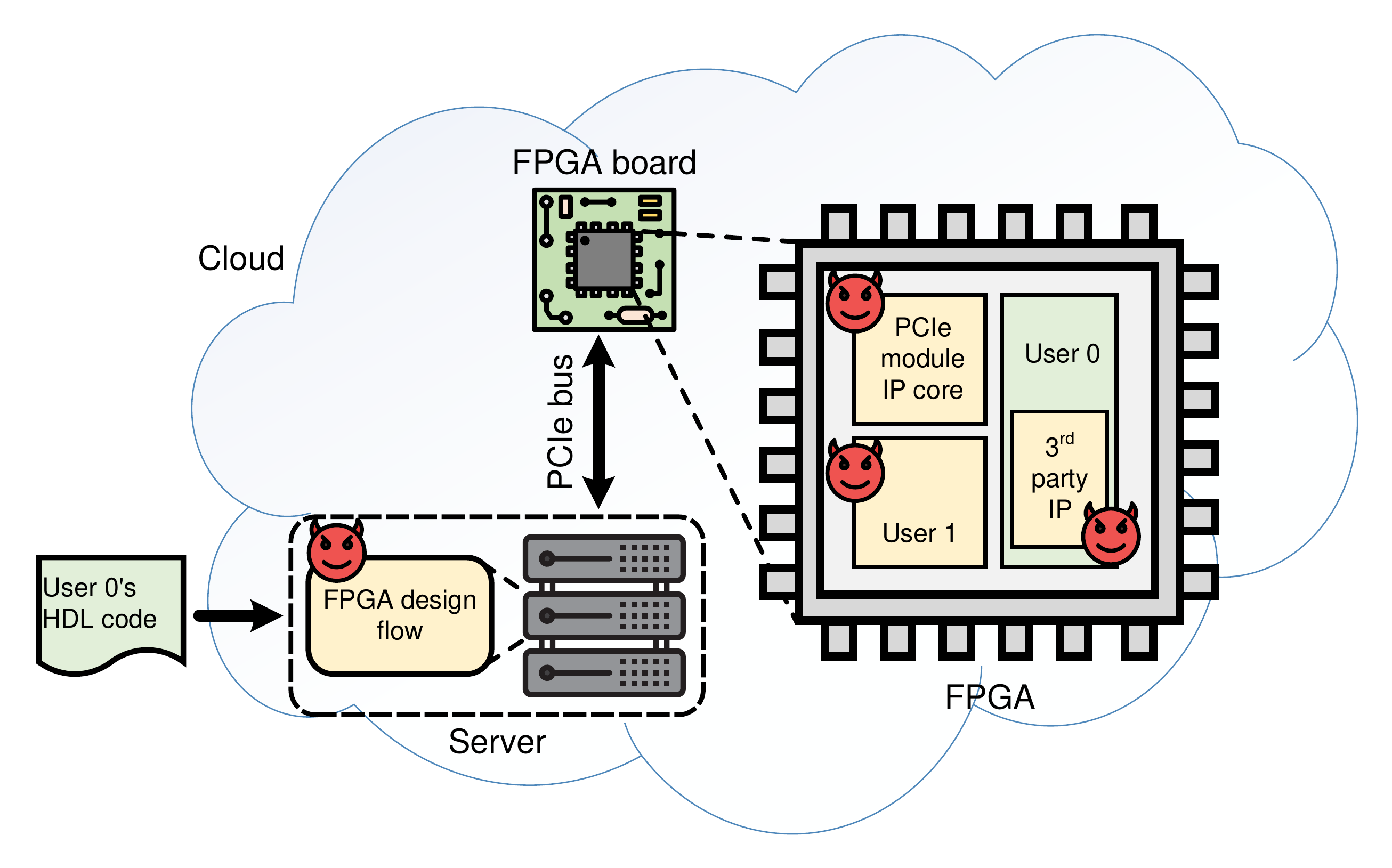}
    \caption{Architecture of an FPGA in the cloud. The four different threat models considered in this paper are (1) malicious cloud providers, (2) malicious co-tenants, (3) malicious IP providers, and (4) malicious FGPA toolchain. These are indicated in the figure by devil icons in the shell (PCIe module and IP core), user 1's logic, 3rd party IP core, and the FPGA design flow, respectively.}
    \label{fig:FPGA_Arch}
\end{figure}

Fig.~\ref{fig:FPGA_Arch} depicts a typical architecture of a cloud platform with FPGAs. FPGA boards are connected with the servers using PCIe wires. PCIe wires are the de facto standard for the communication between a server and the FPGA in commercial FPGA clouds~\cite{aws_slides}. The cloud service providers divide the programmable resources on an FPGA into two parts: (1) the area for implementing the shell,  and (2) the area where users can implement customized logic. The shell includes Peripheral Component Interconnect Express (PCIe) modules, DDR4 DRAM controllers, and control modules, to enable the communication with the servers and DRAM. In Amazon EC2 F1 instances, one out of four DDR4 DRAM controllers is implemented in the shell, and the other three can be implemented in the customized logic~\cite{aws_interface}. Typically, the cloud provider's logic (shell) interacts with user logic via Advanced eXtensible Interface (AXI) protocols~\cite{aws_interface}. On the CPU side, the software development kit provides the application programming interfaces (APIs), so the users with little FPGA experiences can still interact with FPGAs easily~\cite{aws_sdk}. In the modern commercial clouds like Amazon EC2 F1, an FPGA is not allowed to be shared by multiple users due to security concerns~\cite{aws_developer}. However, researchers envision that multi-tenant cloud FPGAs will be realized soon, as it is more cost-effective for both the cloud providers and the users to share resources. Also, the security of multi-tenant cloud FPGAs is an active research area. Thus we will survey recent works on the attacks and defenses on multi-tenant FPGAs as well.

\section{Threat Models}\label{sec:threat}

To understand the possible threats posed to the cloud FPGA users, we categorize the threat models into four types: (1) malicious cloud providers, (2) malicious cloud users/co-tenants, (3) malicious IP providers, and (4) malicious toolchains. Fig.~\ref{fig:FPGA_Arch} illustrates where the threats reside in the architecture of an FPGA cloud.

\vspace{2mm}
\noindent\textbf{Malicious cloud providers.} In the early days of cloud security research, one of the leading security concerns of users was the privacy of their data stored/processed in the clouds~\cite{anthes2010security,brodkin2008gartner}. In traditional threat models of cloud security, the cloud service providers are generally assumed to be untrustworthy, so a user needs to implement his/her security measures to protect him/herself in the clouds. Additionally, the users on the same cloud platform can be a threat to other users, too. However, a malicious cloud model is stricter than the malicious user model because a cloud provider has all the privileges to the platform, including physical access and full control of the computation resources. A typical defense against malicious cloud providers is the use of fully-homomorphic encryption~\cite{gentry2009fully,van2010fully}.

Fully-homomorphic encryption allows computation on encrypted data. Users can encrypt their private data on their computers and send the encrypted data to the cloud. Computation on the encrypted data is performed in the cloud. After the computation, the cloud sends the result back, still in the encrypted form. The user decrypts the encrypted result and gets the result of the computation on his/her private data. As the computation on the cloud is performed completely on encrypted data, the cloud provider is unable to extract any secrets in the user data. Fully-homomorphic encryption is a good way to eliminate the requirement of a trusted cloud. However, it is computationally intensive, and it can take hundreds or even thousands of seconds to complete the bootstrapping operation, which is the most important operation to realize fully-homomorphic encryption schemes~\cite{gentry2011implementing,rohloff2014scalable}. Researchers are developing new encryption constructions and implementations to improve the performance of fully-homomorphic encryption~\cite{brakerski2014efficient,gentry2011implementing,bajpai2014fully}. Also, people have incorporated the homomorphic encryption techniques into secure processor architecture designs as well~\cite{tsoutsos2015heroic}.

\vspace{2mm}
\noindent\textbf{Malicious co-tenants.} Besides the security threats from a malicious cloud provider, threats from malicious users/co-tenants need to be considered. The basic principle of cloud computing is that all the users can dynamically have a share of the large computation resource pool. Due to this, a victim user can be allocated close to a malicious user. Moreover, the victim and the malicious user might even share some computation resources. Although, in general, the computation resources used by different users are logically isolated, the computation resources are likely to be physically connected due to the shared hardware platform. Attackers can leverage such a shared hardware platform to perform a variety of attacks such as side-channel attacks, fault-injection attacks, and establishment of covert channels, which are discussed in the following sections of the paper.

Moreover, the business model of cloud computing pushes the economically-motivated cloud service providers not to act maliciously. Thus, a modern trend in cloud computing research is to consider cloud providers as partners of the users~\cite{aikat2017rethinking}. These providers help protect the security of their customers. For example, cloud providers can apply moving target defense strategies to actively migrate virtual machines within their computing infrastructures~\cite{moon2015nomad}. This bounds the side-channel information leakage as the attacker has to find the new location of the victim before it can carry on the side-channel attack. 

\vspace{2mm}
\noindent\textbf{Malicious IP providers.} The modern hardware design process is very complicated and time-consuming. Practitioners need to integrate 3rd-party intellectual property (3PIP) cores to speed up the development process. This gives attackers a leeway to introduce malicious IPs, and the IPs can be exploited later to leak information, e.g., via covert channels~\cite{giechaskiel2018leaky,giechaskielreading,gnadvoltage,tian2019temporal,giechaskiel2019leakier}. This threat requires the attacker or the attacker's logic to be present in the proximity of the target FPGA fabric. Thus, the attacker can collect leaked information. So, either the cloud provider or a cloud co-tenant has to be malicious as well. However, the vulnerabilities are introduced in the design phase of the victim system, so we consider it as a separate security threat. This security threat is similar to those in the untrusted supply chain of electronics~\cite{rostami2014primer,guin2014counterfeit} and Trojan insertions in pre-silicon hardware~\cite{karri2010trustworthy,haider2017advancing}. Due to this, the usual countermeasures, such as hardware Trojan detection tools~\cite{salmani2016cotd,haider2019advancing}, can be implemented to detect malicious hardware design and IP cores which leak information through their digital output channels. However, novel covert channel communications enabled by a cloud FPGA environment require the immediate attention of the cloud providers and customers. Such covert channel communications are discussed in more detail in Section~\ref{sec:taxonomy}.

\vspace{2mm}
\noindent\textbf{Malicious FPGA tools.} Adversaries can reverse-engineer commercial FPGA design tools and embed malicious functionalities in the toolchain. This way, malicious tools can alter the compiled hardware design. Under this threat model, the adversary can inject Trojans in a design. This maliciously-altered design behaves functionally and formally equivalent to the original design throughout the design flow until the tool writes the design as a bitstream configuration file~\cite{krieg2016malicious}.

\section{Attacks}\label{sec:taxonomy}

Having explained the threat models in the context of cloud FPGAs, we turn our attention to different attacks proposed by researchers. These attacks are grouped according to their threat models.

\subsection{Malicious Cloud Providers}

\vspace{2mm}
\noindent\textbf{Direct sensitive data leakage.} In a cloud without programmable hardware, all the computation and the data are contained in one container (virtual machine). Each container is isolated from another in the hypervisor layer. In the case of a cloud with programmable hardware attached, an attacker with system privilege can tamper with the logic or tap the communication between the FPGA fabric and the processor. This can enable him/her to steal the secret data. In current commercial FPGA-enabled clouds, the FPGA boards connect to the processors via the PCIe protocol. Thus, the cloud provider can intercept the communication between the FPGA boards and the processors with ease.

\vspace{2mm}
\noindent\textbf{Intellectual property theft~\cite{note2008bitstream,benz2012bil}.} The most common use of cloud FPGAs is to implement hardware accelerators for specific computation tasks. The IP of such an accelerator developed and owned by a developer should be protected. Since the developer hands over the bitstream files of the IP cores to the cloud providers, a malicious cloud provider can access the RTL design of the IP core. Bitstream reverse engineering techniques can enable this~\cite{note2008bitstream,benz2012bil}. Thus, a malicious provider can steal the design IP and replicate the accelerator on another FPGA. 

\vspace{2mm}
\noindent\textbf{Tampering with user logic}. A malicious cloud provider can access the user's RTL design. So, during the integration of the user's design with the shell in the cloud FPGA, the providers can introduce malicious modifications in the design. This security threat is also known as hardware Trojans that have been studied for decades~\cite{xiao2016hardware}. On cloud FPGAs, the Trojans can leak sensitive information, which has been protected by other schemes in traditional cloud computing platforms. Also, the Trojans can sometimes be inserted automatically~\cite{jyothi2017taint}. One of the future challenges is to provide a remote attestation feature which allows a remote user to verify the integrity and authenticity of his/her designs in a cloud FPGA. This feature might be similar to the remote attestation provided by Intel SGX~\cite{costan2016intel}.

\subsection{Malicious Co-Tenants} In a multi-tenant FPGA model, many users, including potential adversaries, will share the same FPGA fabric. As the multi-tenant model allows a malicious user to implement his/her design close to a victim, recent research has focused on the security concerns in multi-tenant FPGAs. In particular, remote side-channel attacks~\cite{zhao2018fpga,schellenberg2018inside} and remote fault-injection attacks~\cite{mahmoud2019timing,alam2019ram} have been demonstrated. In this subsection, we survey the existing works on how a malicious co-tenant can use the programmable logic on a cloud FPGA to launch attacks. Note that an adversary can launch these attacks without any administrative privileges.

\begin{figure}[t]
    \centering
    \includegraphics[width=0.6\textwidth]{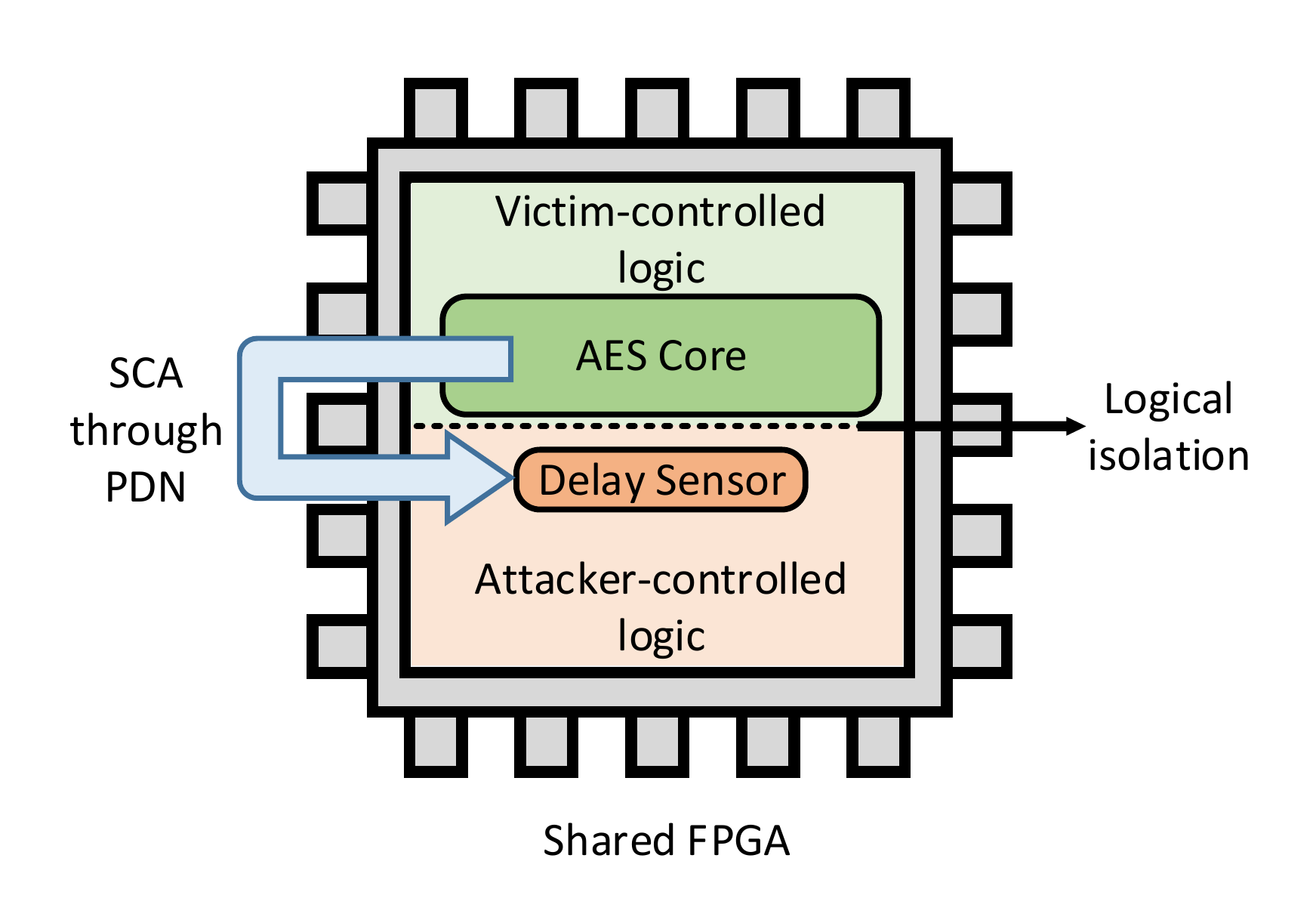}
    \caption{Remote power analysis attack for a multi-tenant FPGA~\cite{schellenberg2018inside}. The side-channel analysis (SCA) is performed through the power distribution network (PDN) in spite of the logical isolation between the victim logic and the sensor.}
    \label{fig:remote_power_analysis}
\end{figure}

\vspace{2mm}
\noindent\textbf{Side-channel attacks.} The attack methods that exfiltrate information that is not leaked through standard digital output channels are called side-channel attacks. Power side-channel~\cite{kocher1999differential,mangard2008power}, timing side-channel~\cite{kocher1996timing,brumley2005remote}, electromagnetic side-channel~\cite{gandolfi2001electromagnetic,carlier2004electromagnetic}, and photonic-emission side-channel~\cite{schlosser2012simple,kramer2013differential} are a few examples of side-channels. An attacker must collect the side-channel information of victim devices in these attacks. Hence, researchers have believed for a long time that the side-channel attacks can be launched only by the attackers with physical access to the devices. However, the ability to program the hardware deployed in the cloud is similar to having physical access to the device. This allows the attackers to monitor the side-channel information remotely in the physical environment, as shown in Fig.~\ref{fig:remote_power_analysis}. The power consumption of a victim logic disturbs the power distribution network on the FPGA, and measuring this disturbance allows the attacker to estimate the power consumption of the victim. Remote power-based side-channel attacks have been demonstrated in the literature~\cite{zhao2018fpga,schellenberg2018inside}. Moreover, crosstalk between FPGA \textit{long} wires (a specific type of routing resource on FPGAs) can also serve as a method to leak information~\cite{ramesh2018fpga}. Since this is an active research area, we provide a detailed survey on side-channel attacks in Section~\ref{sec:case}.

\vspace{2mm}
\noindent\textbf{Fault-injection attacks.} In fault-injection attacks, an attacker injects faults in the execution process of a computation task. Thus, the device produces wrong outputs at the output ports. This problem can have severe implications in a cryptographic system. In such a system, faulty outputs can lead to a successful recovery of the secret key in the system~\cite{guo2015security}. Traditionally, an attacker injects faults by manipulating power or clock signals, or by electromagnetic pulses. These methods require physical access to the target device. However, using FPGAs shared with a victim, an attacker can build an on-chip fault injector and tamper with the computation of the victim.

To demonstrate a successful fault-injection attack on multi-tenant FPGAs, Krautter \textit{et al.} implemented a large number of ROs and program them to oscillate at a very high frequency~\cite{krautter2018fpgahammer}. Because the power distribution network is shared among all tenants on the same FPGA fabric, by toggling the ROs, the attacker can manipulate the propagation delay in the whole chip. Thus, timing violations can occur in the circuit, causing faulty results in the computation. By triggering timing violation on the FPGA, Krautter \textit{et al.} injected faults in an AES process running on the same chip.  Note that this does not require any physical or logical connection to the attacker's circuit~\cite{krautter2018fpgahammer}. Since the high oscillation rate may increase the power consumption drastically, the chip may have to be shut down due to excessive heating. This problem can be addressed, as explained next.

Mahmoud \textit{et al.} improved the fault-injection attack by proposing a delay-sensing circuit. This delay-sensing circuit fine-tuned the parameters of the ROs such that the ROs draw enough power to slow down the target circuit, but not so much that the chip shuts down~\cite{mahmoud2019timing}.

Building ROs is not the only way to generate huge power consumption on an FPGA. Alam \textit{et al.} introduced a new way to inject faults in multi-tenant FPGAs remotely. By repeatedly triggering memory writing collision (writing to the same address simultaneously in a dual-port RAM with opposite values), the attacker can create short circuits in the RAM~\cite{alam2019ram}. This results in massive power consumption in the chip. By exploiting this phenomenon, one can launch a fault-injection attack on an FPGA chip. This attack is stealthier than RO-based attacks because the memory collision can be created during runtime. Such mechanisms, which trigger faults during runtime with unsuspicious circuits, cannot be detected by a bitstream analysis tool, unlike RO based methods, because bitstream analysis tools can detect ROs. Moreover, a dual-port RAM is a common design component in modern hardware system design, which makes the attack more powerful.

\vspace{2mm}
\noindent\textbf{Denial-of-service attacks.} One property of concern for both the cloud providers and the users is the availability of the cloud platform. Denial-of-service (DoS) attackers target the availability of this platform. On an FPGA+CPU heterogeneous cloud, an attacker can launch a remote DoS attack on the FPGA~\cite{gnad2017voltage}. By programming a malicious circuit that switches on and off frequently, a significant voltage drop is created on the FPGA, and the FPGA shuts down to protect itself. An FPGA shut down by voltage emergency requires manual power-cycling of the device.

Matas \textit{et al.} further optimized the RO-based DoS attack by showing how to find the shortest path on an FPGA to form malicious (fastest oscillating) ROs~\cite{matasinvited}. They used GoAHEAD, a tool for implementing partial reconfiguration of FPGAs, to search for the optimal paths~\cite{beckhoff2012go}. The attack, when implemented on a Xilinx Alevo U200 datacenter card with 1.182 million LUTs, can potentially waste over 2kW power, which is way beyond the power budget of any FPGA~\cite{matasinvited}.

\vspace{2mm}
\noindent\textbf{RowHammer attacks.} Interestingly, in an FPGA+CPU heterogeneous system, the FPGA has a unique privilege to access the DRAM without being detected by any monitoring mechanism in the CPU. Also, the FPGA can bypass the cache in the processor and launch a rowhammer attack (i.e., flipping the bits in DRAM by repeated accesses) twice as fast as the traditional rowhammer attack launched by a CPU~\cite{weissman2019jackhammer}. Consequently, the rowhammer from an FPGA to a DRAM can trigger four times as many bit-flips as the CPU initiated attacks. By exploiting this vulnerability, one can tamper with the data and possibly the control flow of the program in the system.

\subsection{Malicious IP Providers}

\begin{figure}[t]
    \centering
    \includegraphics[clip, trim = 0 0.9cm 0 0.7cm,width=0.5\textwidth]{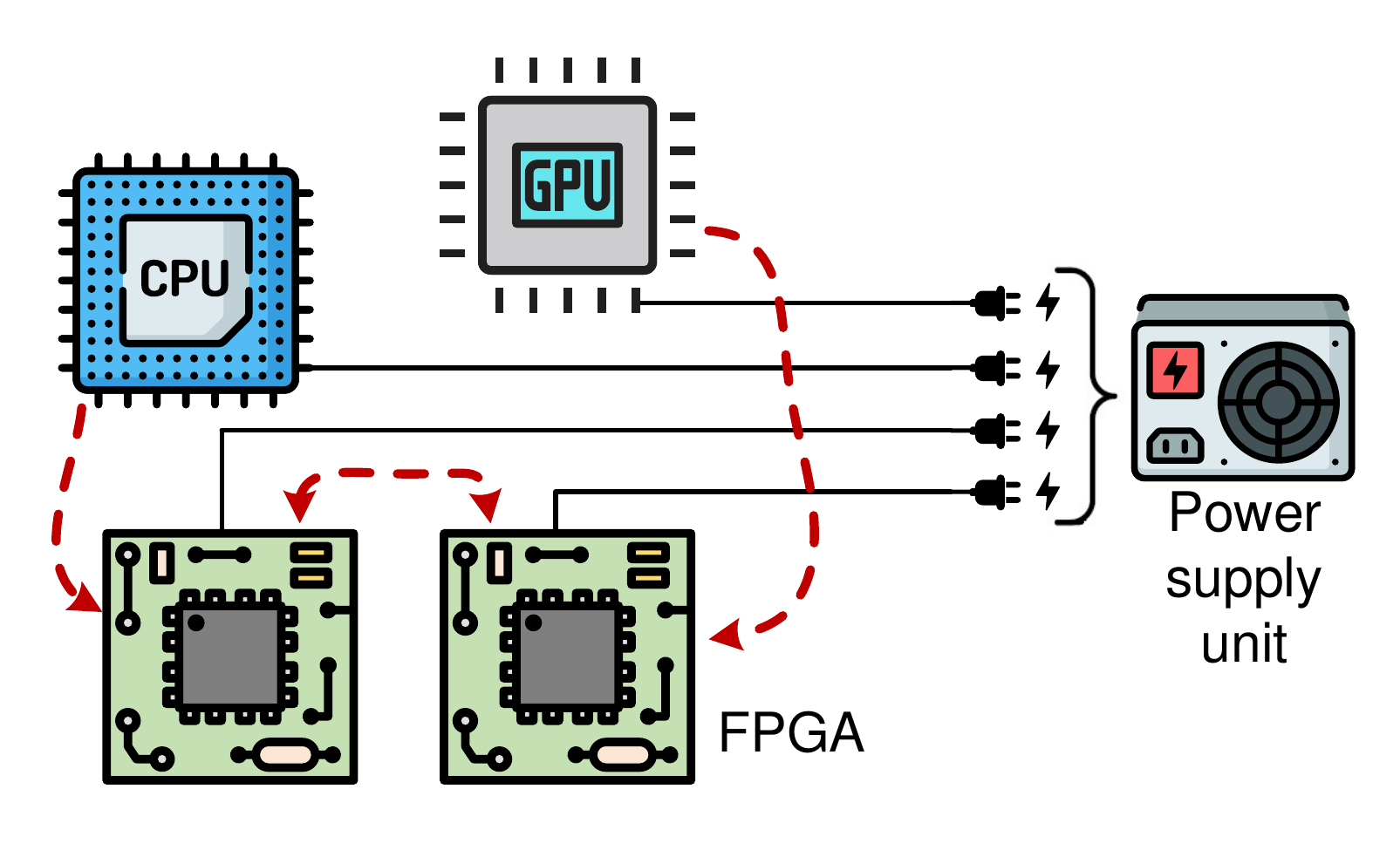}
    \caption{An illustration of power covert channels among CPU, GPU, and FPGAs that share the same power supply unit~\cite{giechaskiel2020capsule}. Note that this attack does not even require the components to share the same power distribution network.}
    \label{fig:CPU_GPU_FPGA_power_attack_color}
\end{figure}

To design a complex modern system, designers usually need to integrate third-party IP cores into their systems. While these third-party IP cores can provide excellent performance and reduce the time to market, they can be security threats to the system. If a malicious IP is introduced in the system, it can leak information or tamper with the computation and the data in the system. Researchers have studied this topic for decades as hardware Trojans. Interested readers can read other survey papers focusing on hardware Trojans~\cite{karri2010trustworthy,xiao2016hardware,bhunia2014hardware}. The unique challenge for exploiting Trojans in the cloud FPGAs is how to leak the information stealthily, which is also called covert channel communication. In the remainder of this subsection, we present how a Trojan circuit can generate side-channel information to send out secrets to a malicious listener. In particular, we show how one transmit over power, crosstalk in \textit{long} wires, and thermal channels. 

\vspace{2mm}
\noindent\textbf{Power covert channels.} The idea of voltage manipulations used in power side-channel attacks can be extended to establish covert channels on multi-tenant FPGAs. An example of this is the work done by Gnad \textit{et al.} in~\cite{gnadvoltage}. They have demonstrated high-speed covert-channel (8MBit/s) communication. The transmitter of the covert channel uses ROs to generate measurable voltage spikes according to the secret data to be transmitted. The receiver, which is another tenant on the same FPGA chip, uses another set of ROs to measure the voltage spikes. The attacker designs both the transmitter and the receiver. This enables the attacker to modulate the transmitted signal leading to robust communication, which can work in the presence of environmental noise introduced by other tenants on the same FPGA fabric.

Establishing such power covert channels can be challenging if the receiver and the transmitter are on separate dies. However, Giechaskiel \textit{et al.} demonstrated such an attack on cloud FPGAs in \cite{giechaskielreading}. They established a power covert channel on cloud FPGAs that are on separate dies. They use Xilinx UltraScale+ FPGAs for this. UltraScale+ FPGAs used by cloud providers like Amazon and Huawei have three distinct dies that are connected and powered through a silicon interposer. Thus, even though the receiver and the transmitter are on separate dies, they still share the same power supply through the silicon interposer. A successful covert channel, operating at more than 4.6Mbps with an accuracy of over 97.6\%, is established in such a setup. Moreover, they showed that the channel is present for all combinations of the three dies as receiver and transmitter.

Sharing a power supply unit in a computing system (e.g., one cloud computing instance), as shown in Fig.~\ref{fig:CPU_GPU_FPGA_power_attack_color}, allows malicious attackers to create a covert channel between FPGA boards, and even from a CPU or a GPU to an FPGA~\cite{giechaskiel2020capsule}. The authors demonstrated that by creating fluctuations in the supply voltage provided by a shared power supply unit, an attacker can send information stealthily to a sink FPGA, which is actively monitoring its voltage by using ROs. The attacker used high power consumption of the source device (FPGA/CPU/GPU) to indicate a logic 1, and low power consumption to represent a logic 0. However, one cannot simply use the absolute RO frequency on the sink FPGA to find out the message sent from the source reliably. This is because the power supply units and voltage regulators on the sink board can tolerate voltage fluctuations to some extent. To solve this problem, Giechaskiel \textit{et al.} implemented stressor ROs on the sink FPGA to drain extra power, so the voltage change in the supply voltage can be more measurable on the sink FPGA. Also, the authors introduced a new metric to detect the power consumption changes on the source device more reliably. According to an evaluation on Artix 7 boards, this covert channel can achieve a bandwidth of 6.1bps with over 90\% accuracy. Similarly, a CPU or a GPU can switch between high and low workload to send bits over the same covert channel to an FPGA.

\begin{figure}[t]
    \centering
    \includegraphics[width=0.6\textwidth]{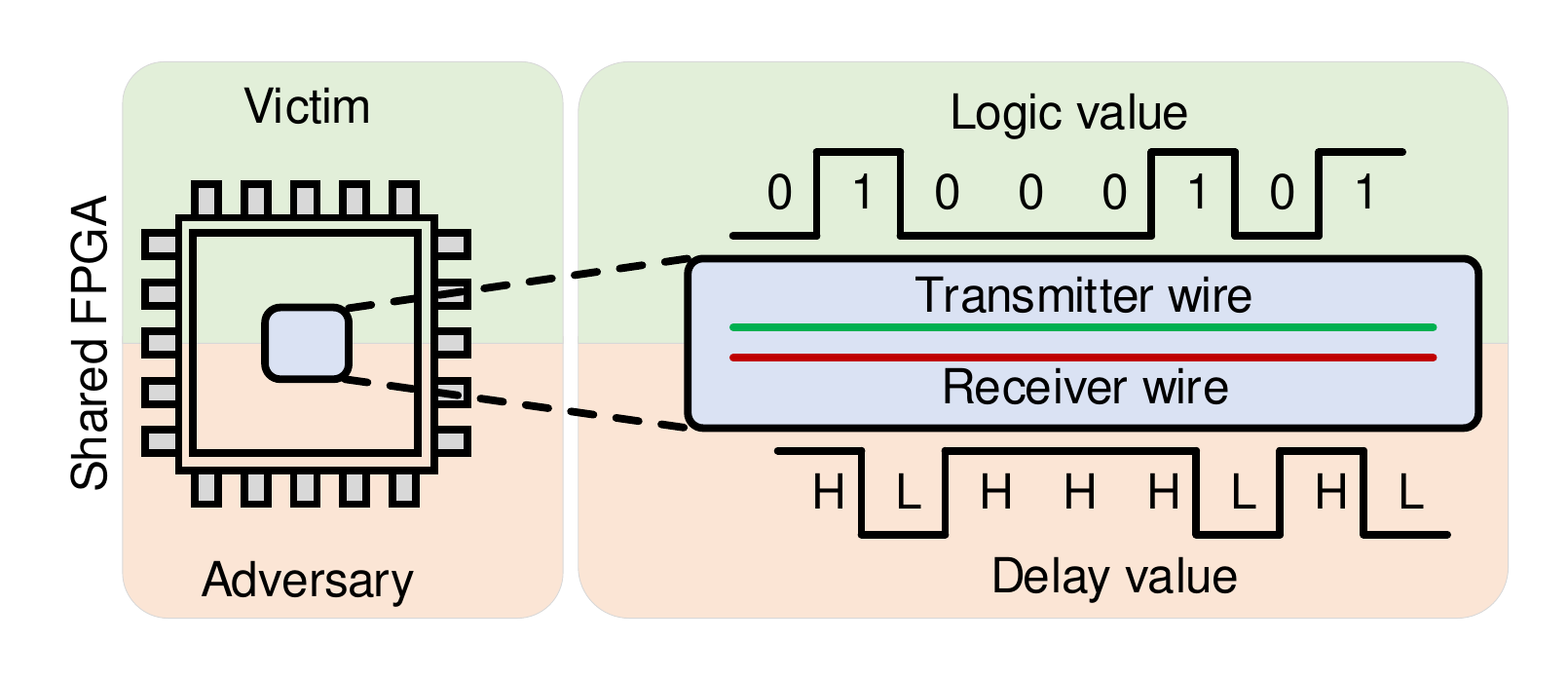}
    \caption{Establishment of a communication channel using \textit{long} wires in Xilinx FPGAs~\cite{giechaskiel2018leaky}. The H and the L on the receiver  stand for high and low delay values respectively.}
    \label{fig:leaky_wires}
\end{figure}

\vspace{2mm}
\noindent\textbf{Crosstalk in \textit{long} wires.} Crosstalk phenomenon in \textit{long} wires can be exploited to launch covert-channel communication as well~\cite{giechaskiel2018leaky}. The attacker is assumed to have a malicious IP core as a part of the victim logic. It is also assumed that the attacker's logic is on the same FPGA fabric and is placed close to the victim's logic. Since the adversary is the designer of the IP core, he/she can define the internal placement and routing of his/her blocks. Thus, the attacker can force his/her cores to use specific routing resources, in particular \textit{long} wires. The attack, illustrated in Fig.~\ref{fig:leaky_wires}, exploits the phenomenon that the delay of FPGA \textit{long} wires depends on the logical state of nearby wires. In particular, when the transmitter wire (the \textit{long} wire in the victim design) carries a logic 1, the delay of the nearby receiving wire (the \textit{long} wire in the attacker's design) is lower than it would be if the transmitter wire carried a logic 0. An RO involving the receiver \textit{long} wire can measure the delay of the receiver wire. This reveals the logic state of the nearby transmitter \textit{long} wire. Thus, a covert-channel is created for attackers to leak sensitive information from a victim hardware design.
This covert channel can work effectively, even in the presence of power and temperature fluctuations. More importantly, the malicious IP core in the victim's logic provides legitimate functionality while acting as a Trojan, and it does not contain additional logic. This makes it challenging to detect such a Trojan using current  Trojan detection tools~\cite{salmani2016cotd,haider2019advancing}. It is worth noting that this attack mechanism does not depend on the rate at which the signals switch. In fact, even when the signal in the transmitter wire is static, the attacker can differentiate between a logic 0/1 on the wire.

Giechaskiel \textit{et al.} extended the idea of~\cite{giechaskiel2018leaky} in~\cite{giechaskiel2019leakier}. They investigated a setup with multiple transmitters and a single receiver in detail. They considered two configurations of the relative placement of the transmitters and the receiver: (1) the two transmitters are on the same side of the receiver (RTT) and (2) the receiver is sandwiched between the transmitters (TRT). In RTT, the transmitter closest to the receiver affects the RO frequency on the receiver. In the TRT configuration, both the transmitters have a roughly equal effect on the RO frequency of the receiver. An attacker can use the TRT configuration to increase the bandwidth or to reduce errors in transmissions.

The effect of crosstalk in \textit{long} wires was characterized in \cite{provelengios2019characterization}. To perform this characterization, the authors proposed a new metric to capture the difference in the periods of the RO for the cases when the transmitter value is 0 and when it is 1. This helps remove the variability in the individual ROs. Experiments showed that this new metric is successful in removing the dependence of the characterization metric on the RO frequency. Hence, it can characterize the leakage more accurately.

A similar characterization effort for cloud FPGAs was made in \cite{giechaskiel2019measuring}. The main challenge in performing this analysis on the cloud FPGAs is the restriction imposed by some cloud providers on the users' designs. Providers like Amazon prohibit combinatorial loops~\cite{AWS_errata}. To bypass this checking on the Amazon cloud, researchers introduced two RO designs in~\cite{giechaskiel2019measuring}. These RO designs are described in Section~\ref{sec:ro_design}.

\vspace{2mm}

\begin{figure}[t]
    \centering
    \includegraphics[width=\textwidth]{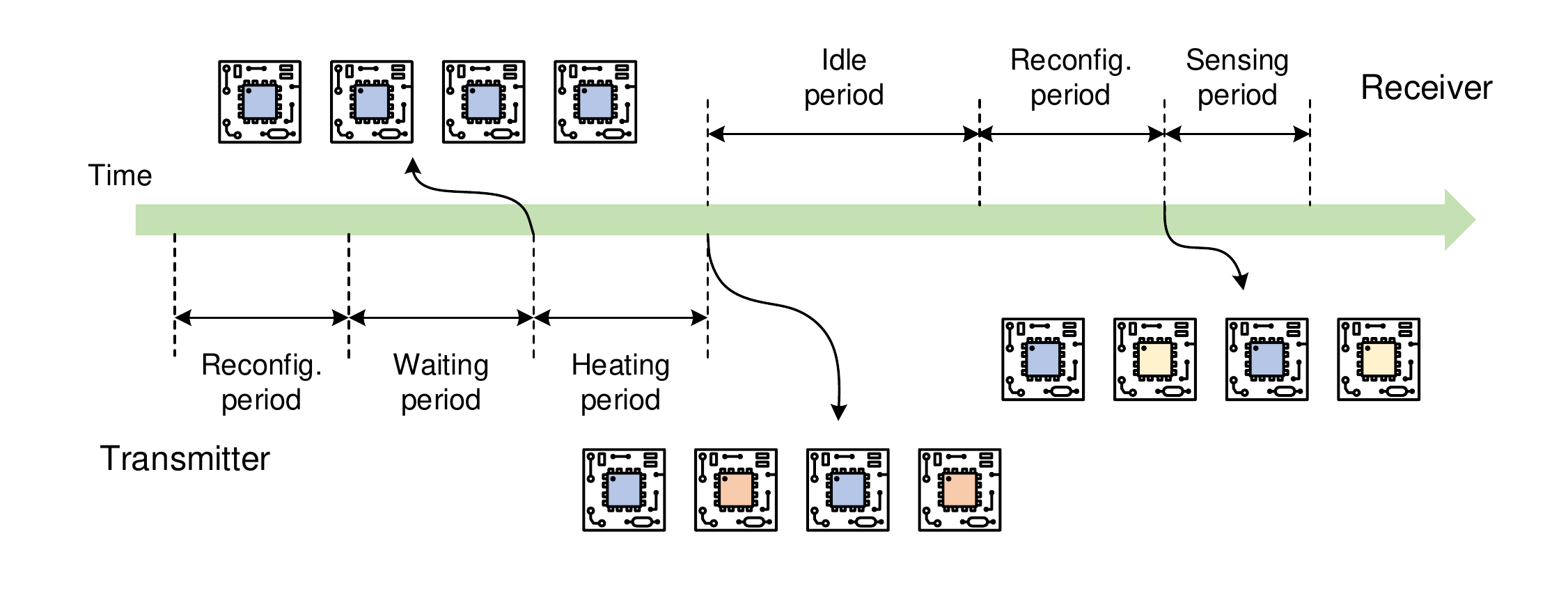}
    \caption{Establishment of thermal covert channel on cloud FPGA~\cite{tian2019temporal}. The transmitter uses 4 FPGAs simultaneously and sends the binary string 0101 in this example. The orange color of the FPGAs after the heating period represents high temperature. The yellow color of the FPGAs after the reconfiguration period on the receiver side represents a temperature higher than the un-heated FPGAs, but lower than the heated FPGAs.}
    \label{fig:Thermal_covert_channel}
\end{figure}

\noindent\textbf{Thermal covert channel.} Most of the covert channels in the literature require the designs of attackers and victims to be present on the same FPGA chip, i.e., a multi-tenant FPGA setup. However, cloud providers have not adopted the multi-tenant FPGA model yet. There exists a covert channel on the cloud FPGAs which does not require a multi-tenant setup. The covert channel described by Tian \textit{et al.} in ~\cite{tian2019temporal} is an example. It exploits the temporal sharing of a single FPGA. This channel can transmit data stealthily on a single-tenant cloud FPGA. The transmitter heats an FPGA by operating many ROs. Then, the transmitter turns off the ROs, leaves the cloud, and the receiver uses the same FPGA. The receiver can measure the temperature of that FPGA with ROs. This is possible because the frequency of an RO depends on the temperature of the FPGA. The bandwidth of such a thermal covert-channel depends on the number of FPGAs used simultaneously. Fig. \ref{fig:Thermal_covert_channel} illustrates how a binary string can be transmitted and received by the temporal sharing of four cloud FPGAs simultaneously. This covert channel was demonstrated on the cloud FPGAs in Texas Advanced Computing Center in \cite{tian2019temporal}.

\subsection{Malicious FPGA Tools}

Users expect that a legitimate vendor provides the FPGA design tools. However, if an attacker can inject malicious functionalities into the toolchain, a malicious bitstream and hence, malicious hardware can be built on the cloud FPGAs~\cite{krieg2016malicious}. The malicious modification in the compiled design does not show up in the output until the bitstream is generated. Hence, the intermediate output files, such as post-place simulation netlists are formally equivalent to the original design. The attacker activates the malicious functionality only when the bitstream is being generated. The malicious FPGA tool first replaces the functional blocks with their malicious counterparts. Then, in the bitstream generation process, the design tool looks for these special malicious LUTs. If the tool finds these malicious LUTs, it reconfigures them to activate the Trojan. The authors demonstrate a privilege escalation attack using this malicious design flow on the free and open-source Lattice iCE40 design flow~\cite{krieg2016malicious}.

\section{Case studies}\label{sec:case}
In this section, we detail the recent research on two popular topics: (1) remote power side-channel attacks on cloud FPGA systems and (2) variants of RO designs on FPGAs, as a building block in many attacks, to bypass the design restrictions enforced by cloud providers. 

\subsection{Remote power side-channel attacks}

Security researchers have studied power side-channel attacks extensively in the past decade~\cite{mangard2008power,kocher1999differential}. An attacker can exploit the fact that the data that the system processes affects the dynamic power consumption of the system~\cite{kocher1999differential}. So, by observing the power consumption of the circuit, the attacker can infer the secret key in the cryptographic hardware. This attack requires side-channel information to be collected from the hardware. Consequently, it was believed that such attacks could be carried out only if the attacker had physical proximity to the target system. However, in the context of cloud FPGAs, a malicious user does not have physical access to the target FPGA. Hence, all previous techniques would not work. This leads to recent works on remote power analysis. We discuss those in this sub-section.

\vspace{2mm}
\noindent\textbf{Threat model.} In general, remote power analysis attacks assume that the adversary's logic and the victim's logic are on the same remote FPGA fabric~\cite{schellenberg2018inside,zhao2018fpga}. So, the adversary has access to some of the LUTs in the remote FPGA. In other words, the attacker can implement his/her logic on some part of the shared multi-tenant remote FPGA. Although currently, the cloud FPGA providers do not allow sharing of an FPGA by multiple users, as explained in Section~\ref{sec:background}, it is envisioned that multi-tenant FPGAs will be realized soon for better efficiency in terms of cost and utilization.

\begin{figure}
    \centering
    \includegraphics[width=0.6\textwidth, trim = 0 0.5cm 0 0, clip ]{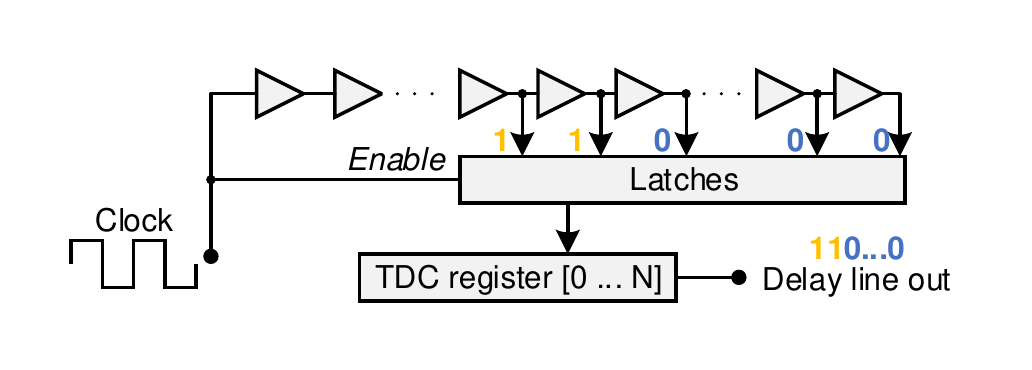}
    \caption{Illustration of a TDC sensor that uses a chain of buffers with latches to measure delay.}
    \label{fig:TDC}
\end{figure}

\vspace{2mm}
\noindent\textbf{Key idea.} To launch a remote power analysis attack, an attacker has to implement a power monitor on the FPGA fabric shared with the victim. For example, the attacker can monitor the power consumption of a victim process by using time-to-digital converter (TDC) sensors. Using the power traces collected by the on-chip power monitors, the attacker can perform a power side-channel attack.

\vspace{2mm}

\noindent\textbf{Attack method.} A key component in the attack is the power distribution network (PDN) on FPGA chips. The PDN handles the distribution of power to all the components on the FPGA~\cite{arabi2007power}. The PDN spans across different abstraction levels, from printed circuit board level to individual transistors on the FPGA. The PDN consists of resistive, capacitive, and inductive elements in the form of a power mesh. The power consumption of an FPGA chip at any instant depends on the logic that is being operated at that time. The changes in logic values affect the voltage and current drawn by the transistors in FPGA. These voltage fluctuations affect the delays of the other logic circuits implemented on the same FPGA due to the shared PDN. Hence, measuring delays in one part of the FPGA reveals information about power consumption in a different part of the FPGA. In particular, the higher the fluctuations in the voltage, the higher is the change in the delays. So, the attacker can monitor the power fluctuations on the FPGA by implementing appropriate delay sensors.

To this end, the attacker can implement a TDC, illustrated in Fig.~\ref{fig:TDC}, on the shared FPGA as a delay sensor~\cite{schellenberg2018inside}. As the delays of the buffers in the TDC depend on the supply voltage, the change in delays can be monitored as a proxy for voltage fluctuations. When a victim process becomes active in a different region of a multi-tenant FPGA, it disturbs the PDN. This results in a change in the delay values of the TDC sensor. Thus, the attacker can create a mapping between the power traces and the delay values. This mapping can then be used to perform a standard Correlation Power Analysis (CPA) attack. Such an attack was demonstrated in~\cite{schellenberg2018inside}.
The proof of concept for this attack was demonstrated on a victim AES core operating at 24MHz on a Xilinx Spartan-6 FPGA. Two scenarios were considered: (1) when the sensor is placed close to the victim AES logic, with a gap of just 4 FPGA slices, and (2) when the sensor is placed far from the AES core. In both cases, the attacker can recover the AES key.

\vspace{2mm}
\noindent\textbf{Alternate power sensors.} As a different approach, Zhao \textit{et al.} used ROs as power sensors to monitor the power consumption on the FPGA~\cite{zhao2018fpga}. They translated the frequencies of  ROs into the power traces based on a linear relationship. A remote power side-channel attack was shown to be successful using this RO sensor setup.

\vspace{2mm}
\noindent\textbf{Attacking the processor system.} In an FPGA+CPU heterogeneous chip, like a Xilinx Zynq system, an ARM processor system (PS) shares the PDN with the FPGA fabric (programmable logic or PL). Zhao \textit{et al.} demonstrated an attack that uses the PL to monitor the power consumption of the PS~\cite{zhao2018fpga}. By doing so, they recovered the control flow of the program in the PS. This vulnerability made a simple power analysis on RSA possible. Similarly, an FPGA-to-processor correlation power analysis has been demonstrated in~\cite{gravellier2019remote}. The authors used a TDC on the FPGA to measure the power traces of the processor. Using that, they attacked an AES core running on the processor with 111k to 127k power traces.

\vspace{2mm}
\noindent\textbf{Cross-chip attacks.} Using the remote power side-channel attack, an attacker can not only attack the victim who is on the same chip as the attacker, but he/she can also launch a cross-chip attack. This cross-chip attack works as long as two FPGA chips are sharing the same power supply on the same board~\cite{schellenberg2018remote}. Due to the victim being on a separate chip, the attack is more challenging. The number of traces required for this attack on AES is 40$\times$ the number of traces required for the attack in~\cite{schellenberg2018inside}.

\vspace{2mm}
\noindent\textbf{Experiments on Amazon clouds.} In DATE'20, Glamocanin \textit{et al.} published their results on launching remote power side-channel attacks on AWS EC2 F1 instances~\cite{glamocanin2020are}. They chose to use TDC sensors for measuring power consumption on a cloud FPGA, and the results showed that they could successfully break the secret keys of all 16 bytes of an open-source AES-128 core with $5 \times 10^5$ traces. This result validated the feasibility of remote power side-channel attacks on a commercial cloud platform, so this research area raises serious concerns.

\vspace{2mm}
\noindent\textbf{\textit{Long} wire leakage.} Ramesh \textit{et al.} showed that it is possible to exploit the crosstalk phenomenon in \textit{long} wires to extract a secret from a victim logic passively~\cite{ramesh2018fpga}. In the attack, the authors targeted an automatically placed-and-routed AES core. They identified a vulnerable \textit{long} wire in the victim design, which carries secret information. This vulnerable \textit{long} wire would act as the transmitter. The attacker is assumed to manually place-and-route a \textit{long} wire in the receiver RO such that it is adjacent to the vulnerable (i.e., transmitter) \textit{long} wire. After that, a side-channel attack based on the \textit{long} wire leakage is conducted successfully. In this attack, the attacker does not need to modify the routing constraints of the victim logic. However, a successful attack relies on the fact that the FPGA tool has created a vulnerable design.

\begin{figure}[t!]
    \centering
    \includegraphics[width=0.6\textwidth]{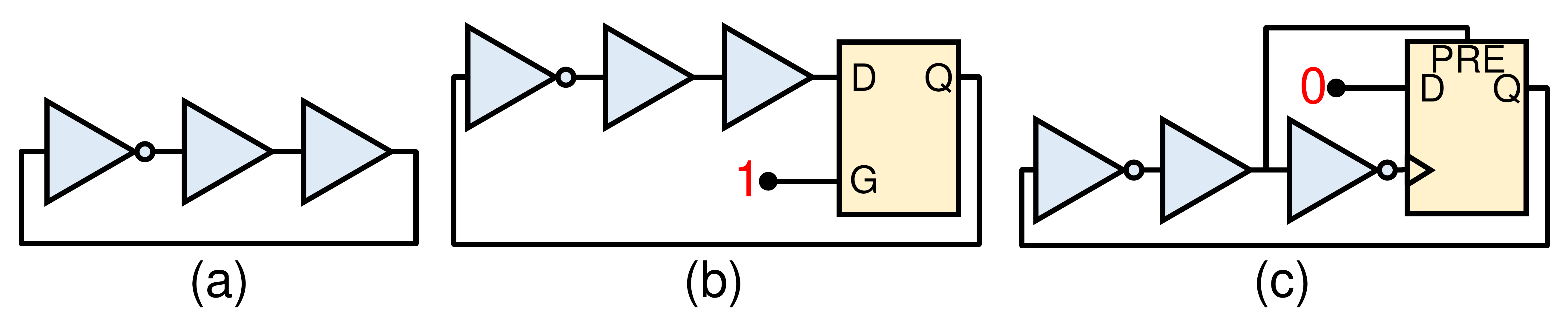}
    \caption{Classical RO design (a), and its variants (b) and (c) \cite{giechaskiel2019measuring}.}
    \label{fig:RO_designs}
\end{figure}

\begin{figure}
    \centering
    \includegraphics[width=0.6\textwidth]{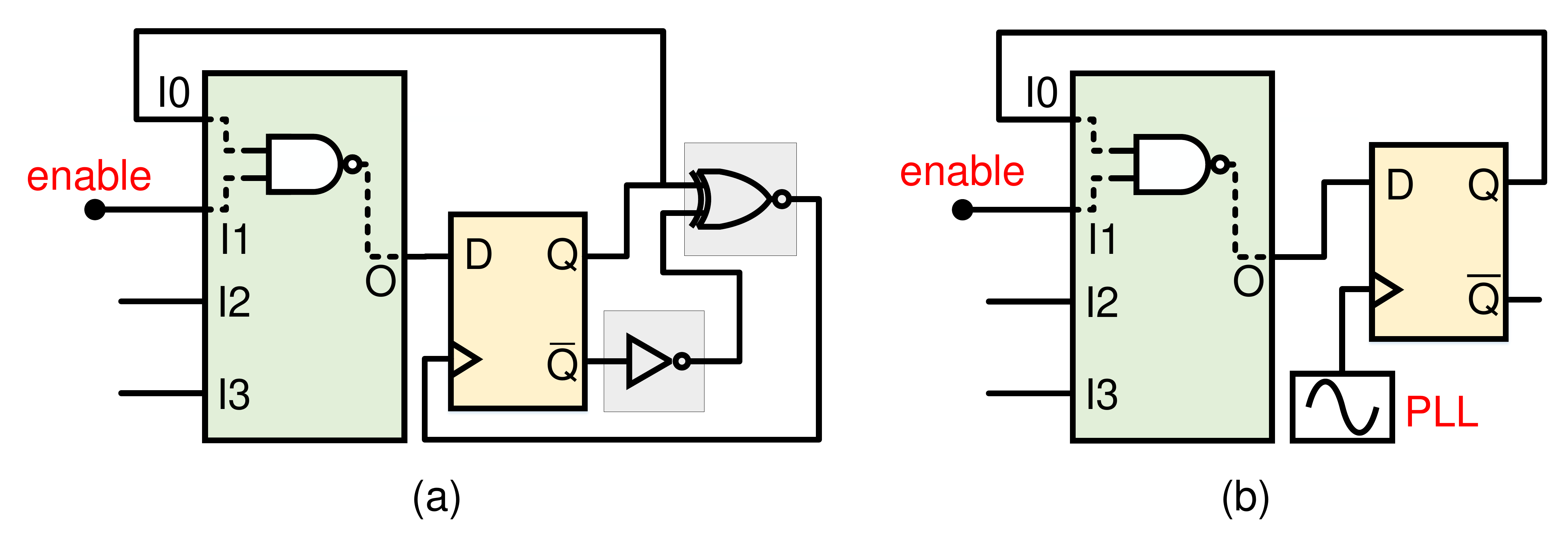}
    \caption{High switching activity components~\cite{krautter2019mitigating}: These designs are potential replacements for classical ROs in fault injection attacks since they incur high power consumption as well. The inverter and the XNOR gates in the grey boxes in (a) are preserved to add delay and generate clock glitches for the flip-flop. The circuit in (b) performs switching by using a high-frequency signal from the Phase-locked loop (PLL) as the clock of the flip-flop.}
    \label{fig:Other_RO_designs}
\end{figure}

\subsection{Ring Oscillator Designs and their Variants.}\label{sec:ro_design}
ROs are a crucial component in a variety of attacks~\cite{zhao2018fpga,ramesh2018fpga,giechaskiel2019leakier,giechaskiel2018leaky,provelengios2019characterization}. So, we survey different RO designs and their variants that exist in the current literature.

Typically, an RO is composed of a self-looped chain of an odd number of inverters. Each inverter can be instantiated on a LUT in an FPGA, as shown in Fig ~\ref{fig:RO_designs}. The reason why a variety of attacks rely on ROs is the sensitivity of the RO frequency to voltage and temperature fluctuations. However, due to the possible use of ROs in attacks, AWS implements a netlist checking tool and blocks users from implementing combinational loops (i.e., typical ROs) on their FPGAs. So, the basic design of an RO has been extended to designs that use a latch or a flip-flop as a transparent component like a buffer~\cite{giechaskiel2019measuring}. Such designs are illustrated in Fig.~\ref{fig:RO_designs} (b) and (c). An advantage of having sequential elements in ROs is that it can fool a bitstream or netlist checking tool into believing that the design does not contain a combinational loop.  So, this technique can be used to hide the existence of combinational ROs in the design from the checking tools, as demonstrated in~\cite{giechaskiel2019measuring}.

The above-mentioned RO design variants can be used to replace classical ROs in both side-channel and fault injection attacks. Researchers have come up with different designs that can incur high switching activity, and thus imply a high power consumption~\cite{krautter2019mitigating}. These designs are shown in Fig.~\ref{fig:Other_RO_designs}, and they can be used in fault injection attacks as power wasters. These designs have the oscillation property of an RO, but they are sequential circuits since a flip-flop is involved in the loop. The main design principle of high switching activity components is to toggle values as fast as possible. In the first design in Fig.~\ref{fig:Other_RO_designs}, a self-clocked flip-flop is used, and the clock signal is generated by an XNOR gate which generates glitches. The second design in Fig.~\ref{fig:Other_RO_designs} uses a phase-locked loop (PLL) at its highest frequency to generate a high-speed clock. On an iCE40-HX8K FPGA, the current generated by 6000 combinational ROs and 6000 PLL based sequential oscillators in Fig.~\ref{fig:Other_RO_designs} (b) is measured to be 291.3~mA, and 240.9~mA, respectively~\cite{krautter2019mitigating}. Note that the sequential oscillator cannot consume as much power as combinational ROs because it cannot run as fast as a combinational RO. Still, the sequential design is powerful enough to launch fault attacks on FPGAs. 
\section{Countermeasures}\label{sec:counter}
Several researchers have proposed methods to counter the attacks mentioned in the previous sections~\cite{gnad2018checking,krautter2019mitigating,provelengios2019characterizing,krautter2019active}. These methods can be classified broadly into two categories: defenses implemented by a tenant and defenses implemented by a cloud provider.

\subsection{Untrusted Clouds}

\vspace{2mm}
\noindent\textbf{Bitstream encryption.} One way to prevent bitstream reverse engineering is to encrypt the bitstream~\cite{bitstream_enc}. The encrypted bitstream is decrypted only on the FPGA. Many commercial FPGAs support this feature. However, so far, to the best of our knowledge, commercial heterogeneous cloud providers (e.g., Amazon EC2 F1) require users to submit RTL designs in plaintext~\cite{aws_hdk}. The platform provider then integrates such a design with the shell (i.e., the PCIe modules and the control modules for communicating with the servers). Before the generated bitstream file is loaded onto an FPGA, the cloud platform providers check the design for prohibited design patterns like combinational loops. Thus, commercial clouds do not support bitstream encryption. Apart from this method, there are no perfect solutions to prevent bitstream reverse engineering.

Following the line of bitstream encryption, Bag et al. proposed a key management system to manage the bitstream decryption process on FPGA~\cite{bag2018cryptographically}. They combined the concept of ``bring your own key'' (BYOK) and key aggregate cryptosystem~\cite{patranabis2016provably}. This way, the tenants can use their secret keys for encrypting their bitstreams locally and securely transfer the keys to the cloud FPGAs for decryption. A master public key provided by the FPGA vendor encrypts the encryption keys used by each tenant. The master private key is embedded in the FPGAs and can decrypt the secret keys of the users. These decrypted keys are then used for decrypting the corresponding encrypted bitstream files. A malicious cloud provider does not have access to the secret keys without using the FPGA, because the decryption for the individual secret keys occurs in the FPGA.

\vspace{2mm}
\noindent\textbf{IP watermarking.} 
IP watermarking is a technique that adds special modules into a hardware design (IP core). It should be difficult for an attacker to detect and remove the embedded watermarks. Moreover, the watermarks should be embedded such that the owner can prove the ownership of the design when an IP dispute occurs. For example, \cite{lach1998signature} encodes watermarks in the unused LUTs in an FPGA, and the design is placed and routed around the watermark. As the location of the watermark is known only to the designers, an attacker cannot detect which LUTs contain the watermark bits. Some watermarking techniques incur extra area overhead to the design~\cite{lach1998signature}. However, zero-overhead watermarking schemes exist. For example, one can modify the delay of non-critical paths in the hardware design through routing, such that the delay information can be considered as the unique watermark of the IP core~\cite{jain2003zero}. Since this technique does not add any hardware components, it has no additional hardware area overhead. 

\vspace{2mm}
\noindent\textbf{Traditional side-channel attacks and fault-injection protections.} When the cloud provider is untrustworthy, it is difficult for the user to protect him/herself against the cloud and the other malicious tenants. This is because the user has no control over and no idea about who and what will be sharing their FPGA, and he/she has to assume everyone else is potentially malicious to him/her. One conservative method to design a secure system against side-channel attacks and fault injection attacks is to follow the traditional security practices and assume that the attackers have physical access to the device. Researchers have studied side-channel attacks and fault-injection attacks for decades. We generally understand how to secure a hardware design against side-channel attacks (e.g., using masking or hiding principles~\cite{mangard2008power}) and fault-injections (e.g., using fault-injection defections~\cite{guo2015security,endo2012efficient}). However, to defend against such a strong physical adversary, the area and the performance overheads of the design are typically very high. Thus, these kinds of countermeasures may be an overkill for the scenario of cloud FPGAs, where the attackers may not be able to precisely measure the side-channel information and inject faults using the on-board malicious circuits. We are still seeking more efficient ways to secure the designs on cloud FPGAs, while not relying on the trustworthiness of the cloud providers.

\subsection{Trusted Clouds}

\vspace{2mm}
\noindent\textbf{Bitstream checkers.} If a cloud provider is trustworthy, a non-malicious user can be at an advantage, since the cloud provider can use the unique insights it observes from the whole platform to implement more powerful countermeasures. For example, the provider can implement a bitstream checker before programming an FPGA. This bitstream checker can identify malicious circuit structures and raise a red flag if any such structures are found. The provider can use this method of checking the bitstream to defend against side-channel attacks and fault-injection attacks. For instance, AWS does not allow a user to deploy ROs and combinational loops on their FPGAs~\cite{AWS_errata}.

Being one of the pioneers in this research direction, Gnad \textit{et al.} built the first bitstream checker called \textit{FPGA antivirus}~\cite{gnad2018checking}. It checks for the known patterns of malicious circuits. They identified one typical pattern for fault injectors and two different patterns for the sensor designs used in side-channel attacks. 
Previous research has demonstrated that a fault can be injected by voltage or current fluctuations on the FPGA fabric~\cite{krautter2018fpgahammer,mahmoud2019timing}. 

Gnad \textit{et al.} identified two characteristics of an on-board fault injector: (1) a large number of combinational loops, where the "largeness" threshold is determined empirically; (2)  a common input to these combinational loops to synchronously toggle the behavior of the loop. The rationals behind these two characteristics are: (1) to launch a fault-injection attack, the attacker has to be able to control a circuit that can incur high switching activities and extremely high power consumption; (2) the large circuit has to be synchronically controlled, otherwise, the large power consumption will be flattened, so the current will not surge beyond the current limit of the FPGA. 

There are two ways to launch power side-channel attacks on multi-tenant FPGAs. One uses ROs as sensors, and the other uses a time-to-digital converter (TDC) as a sensor~\cite{schellenberg2018inside}. Gnad \textit{et al.} suggested two types of characteristics to identify the on-board sensors from bitstream files. To identify the RO sensors, they looked for combinational loops with output ports. To identify the TDC sensors, they checked for timing violations on every wire. These checks for timing violations apply to other sensors that exploit timing violations as well. Even if a single bit is unstable during a voltage fluctuation, an attacker can use that bit to measure the delay on the FPGA.

Based on the patterns of malicious circuits mentioned above, one can catch all malicious circuits known by then. However, as the authors in~\cite{gnad2018checking} have noted, potentially, there are more ways to launch side-channel and fault attacks on-board. So, this pattern-based approach needs to be updated frequently to catch up with new attacks~\cite{gnad2018checking}.

Soon after the original \textit{FPGA antivirus} was built, Krautter \textit{et al.} proposed new attack variants that can evade the detection of the original \textit{FPGA antivirus}~\cite{krautter2019mitigating}. These new attacks rely on novel RO structures designed using sequential circuits~\cite{sugawara2019oscillator}, so analysis tools that look only for combinational loops cannot detect such RO structures. To keep \textit{FPGA antivirus} up to date, Krautter \textit{et al.} reformulated the necessary characteristics of potentially malicious circuits that can launch fault-injection attacks and side-channel attacks~\cite{krautter2019mitigating}. They concluded that to detect side-channel attacks, a bitstream checker needs to look for sensors that can detect delay changes on board. Thus, they identified three unique patterns of malicious circuits for side-channel attacks: (1) paths with timing violations; (2) unusual data to clock connections; and (3) ROs. Likewise, they also summarized the characteristics of malicious fault-injectors: (1) high current variation runtime behaviors; (2) a large number of synchronized elements; (3) hardware primitives which can oscillate, e.g., ROs. They combine static analysis and dynamic analysis in the new \textit{FPGA antivirus} design. The static analysis checks the structural properties of the design, and the dynamic analysis looks for possible timing violation and estimates the power consumption based on real or random input stimuli. The updated \textit{FPGA antivirus}, when evaluated on Lattice FPGA, can detect all known malicious circuits that can inject faults or observe side-channel information~\cite{krautter2019mitigating}.

Similarly, Matas \textit{et al.} presented another bitstream checker called \textit{FPGADefender}~\cite{matasinvited}. \textit{FPGADefender} relies on static analysis of bitstream. It looks for structural signatures of malicious circuits. This includes combinational loops with/without transparent latches, short circuits, antennas, large fan-outs, disallowed port and path usage, and latches. 

To counter the security threats from malicious FPGA tools, the authors in \cite{krieg2016malicious} suggested using equivalence checking. This equivalence checking would reveal manipulations in the bitstream file. However, as bitstream formats are not publicly documented, it is hard for third-party verification tool vendors to offer solutions that prove equivalence. Thus, a call for open and publicly documented bitstream formats was made in \cite{krieg2016malicious}. However, even if the bitstream formats are publicly documented, as the design complexity increases, more sophisticated equivalence checking methods are needed.

One drawback of the bitstream checking approach is that the users need to rely on the cloud providers to check the bitstreams. If the providers can check the bitstreams, they can reverse engineer the designs submitted by users. This forces the users to trust the cloud providers.

\vspace{2mm}
\noindent\textbf{Access control.} On a trustworthy cloud, the provider can implement several defenses. For example, the provider can enforce proper access control policies between a processor and its hardware peripherals on an FPGA. Elnaggar \textit{et al.} introduced a new security threat on a partially reconfigurable FPGA. They assume that the reconfiguration manager and the internal/processor configuration access port (ICAP/PCAP) are compromised~\cite{elnaggar2019multi}. This leads to three new attack scenarios: (1) malicious or unregistered bitstream files can be loaded to the FPGAs; (2) unauthorized software can access user logic; (3) attackers can redirect messages between a software application and its custom logic to a malicious application.

To defend against a compromised reconfiguration manager and ICAP/PCAP, \cite{elnaggar2019multi} suggested adding a secure authentication module (SAM), a task/application loading module, and a secure task database into the system. The SAM distributes a shared secret key to users and asks them to embed the key in their applications and hardware modules. By actively running a challenge-response protocol between a software application and its hardware task modules, the SAM can verify the authenticity of applications and tasks, and the applications and tasks can mutually verify the authenticity of each other. Also, the issue of running an unregistered hardware task (or task hiding in~\cite{elnaggar2019multi}) can be remedied by introducing a secure task database and enforcing every authorized user to register her tasks in the database.

The authors in \cite{hategekimana2018secure,yazdanshenas2018improving} introduced access control policies or an encryption core to secure the communication between the processors and the hardware accelerators on the FPGA. However, they overlook the threat of a malicious co-tenant on the same FPGA who can launch side-channel ~\cite{schellenberg2018inside,zhao2018fpga} and fault-injection attacks~\cite{krautter2018fpgahammer}.

\begin{figure}
    \centering
    \includegraphics[width=0.6\textwidth, clip, trim = 0 1cm 0 0]{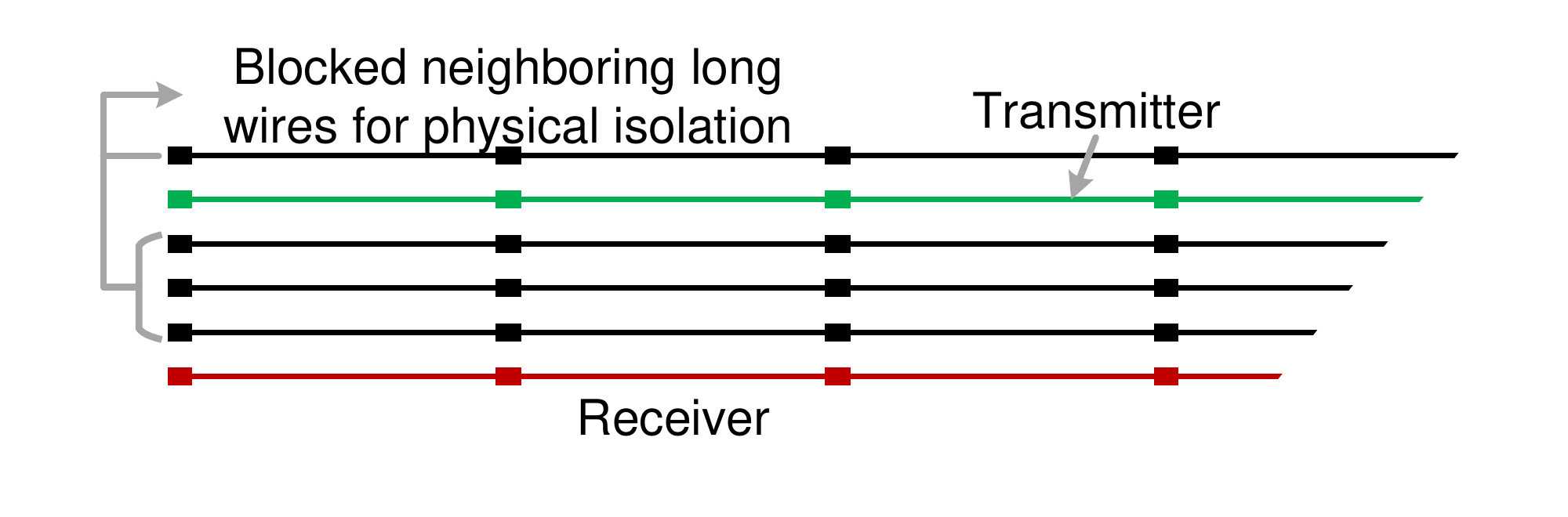}
    \caption{Physical isolation by blocking the adjacent wires upto a distance of 3 from the potential transmitter wire~\cite{giechaskiel2018leaky}.}
    \label{fig:physical_isolation}
\end{figure}

\vspace{2mm}
\noindent\textbf{Physical isolation.} The most effective method to prevent \textit{long} wire crosstalk effect-based attacks is physical isolation. Huffmire \textit{et al.} proposed the concept of physical moats on FPGAs to isolate the hardware cores of different users~\cite{huffmire2007moats}. The moats are implemented using disabled switchboxes (SBs) surrounding each hardware core. The width of the moats (the number of disabled SBs) depends on the number of SBs that can be skipped in routing a \textit{long} wire on the FPGA. In practice, as it is suggested in~\cite{giechaskiel2018leaky}, minimum width of three, as depicted in Fig.~\ref{fig:physical_isolation}, should be enforced to minimize \textit{long} wire crosstalk. Also, if the location of the attacker on the FPGA can be known and constrained, then one can use the FPGA Trust Zone technique proposed in~\cite{jyothi2016fpga} to avoid the FPGA regions adjacent to the attacker.

As the first step in automating secure routing to mitigate the crosstalk effect in \textit{long} wires, Seifoori \textit{et al.} extended an existing open-source FPGA routing tool, PathFinder, to build a routing tool to prevent crosstalk-based side-channel leakage~\cite{seifoori2020closing}. To use the proposed tool, users need to annotate the trusted IP cores and sensitive FPGA nets (e.g., the ones carrying secret keys). In the proposed tool, Seifoori \textit{et al.} proposed four routing strategies. (1) Block-2NN, meaning that no net is allowed to use the nearest and the second nearest \textit{long} wires of a sensitive net; (2) Block-NN, meaning the nearest \textit{long} wires of a sensitive net will not be occupied; (3) Block-Untrusted, no nets from an untrusted IP module can be allocated adjacent to a sensitive net; (4) Lock-NN, meaning the nearest \textit{long} wires of a sensitive net can only be occupied by the nets originating from the same module as the sensitive net. Using the Verilog-to-Routing benchmark~\cite{rose2012vtr}, they found that the four strategies incur 1.91\% to 7.69\% overhead in channel width on average with respect to the baseline, and secure routing introduces 0.12\% to 1.18\% increase in critical path delay on average.

An automatic hardware isolation framework, HILL, was presented in~\cite{luo2019hill}. Given a list of security-critical net names in a design by the designers, HILL can automatically generate a constraint file for an FPGA tool (e.g., Vivado or Xilinx ISE) to place the critical instances (e.g., LUTs) in the middle of the hardware design and route all the other instances in a spiral manner around the critical instances. Thus, the critical instances are protected by the non-critical instances in the protected design from an attacker who is placed outside of the design. If the width of the surrounding non-critical instances is sufficient, the attacker cannot exploit the crosstalk phenomenon to leak information from critical instances. Moreover, for some \textit{long} wires which cannot be placed in the middle of a design, like IO buffers, the authors suggested adding two dummy \textit{long} wires to be adjacent to the vulnerable \textit{long} wires to obfuscate the observation of an attacker. 

\vspace{2mm}
\noindent\textbf{Runtime monitors.} Runtime monitoring is a general defense methodology that deploys performance monitors on the FPGAs to detect suspicious behaviors during runtime. Without compromising the users' privacy in the hardware designs, the cloud provider can deploy runtime monitors on the FPGAs to monitor the running status of the FPGAs.  For example, ROs were proposed to check the delay variations~\cite{zick2012low,boemo1997thermal}, and TDC was introduced in~\cite{zick2013sensing} for sensing nanosecond-scale voltage variations. These sensor designs were all proposed before the first remote side-channel attacks on FPGAs were conducted. 

To take one step further in this direction, Provelengios \textit{et al.} characterized the behavior of a power distribution network on an FPGA under a power-based fault-injection attack~\cite{provelengios2019characterizing}. In particular, they studied the geographic distribution of a voltage drop around a power waster (the source of attacks, e.g., ROs). Essentially, the closer a circuit is to the power waster, the more voltage drop it experiences. Based on this finding, one can build a distributed voltage monitor network on an FPGA to identify, in real-time, the location of the malicious circuit. Then, the cloud provider can move the suspicious user's circuit to another single-tenant FPGA or  remove the user from the cloud. 

One drawback of using a runtime monitoring approach is that it can only detect active attacks like fault-injections. The existing runtime monitors can not check whether there exists an attacker in the chip who is monitoring the system quietly.

\begin{figure}
    \centering
    \includegraphics[width=0.55\textwidth]{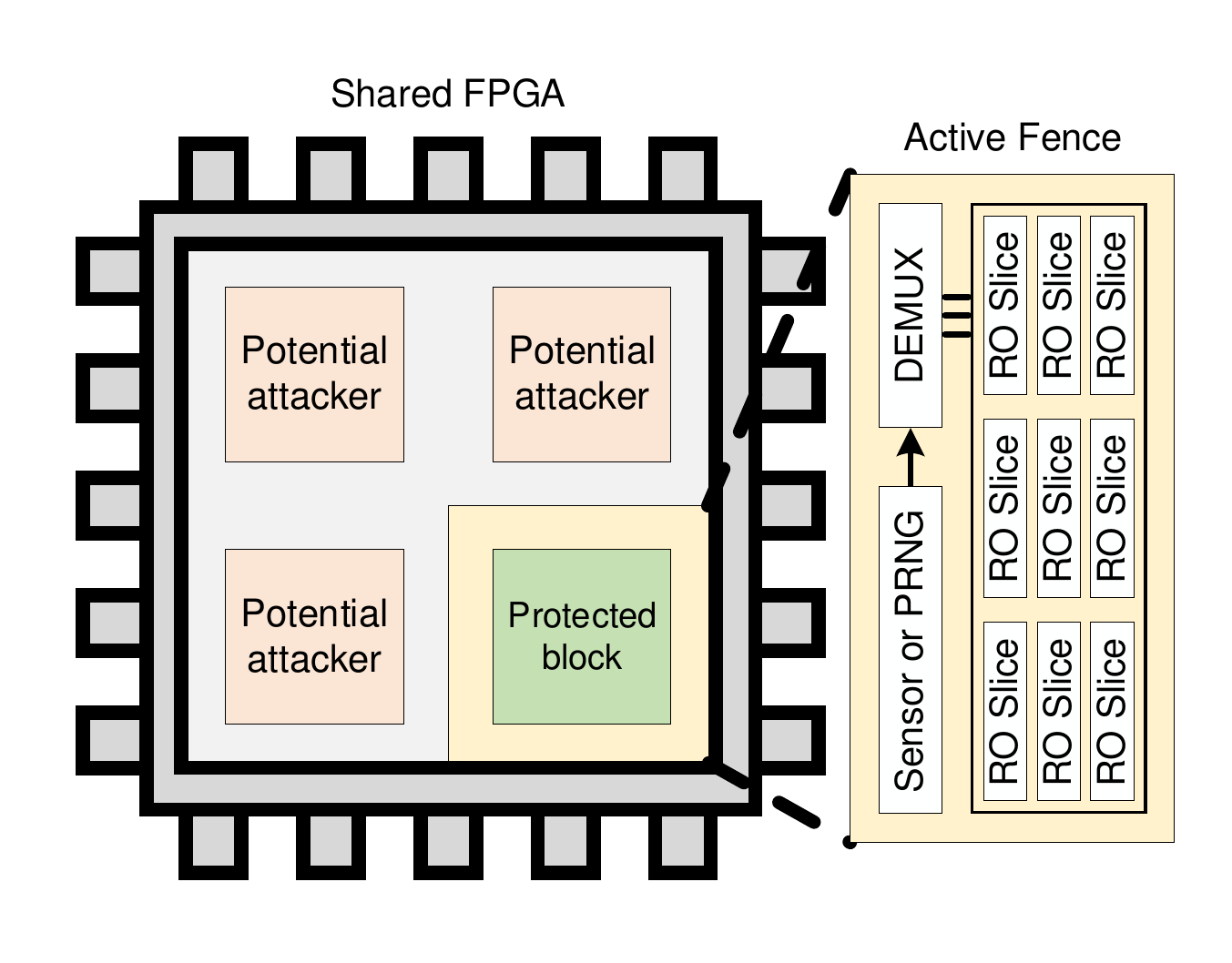}
    \caption{Active injection of noise into power traces by activating an appropriate number of ROs~\cite{krautter2019active}.}
    \label{fig:active_fences}
\end{figure}

\vspace{2mm}
\noindent\textbf{Active defenses.} To counter the passive attacks, like side-channel attacks, a user can actively inject noise into the power traces~\cite{krautter2019active}. This paper introduced active fences between the victim circuits and power side-channel attack circuits, as shown in Fig.~\ref{fig:active_fences}. The fences are composed of ROs, and these ROs can be activated by two approaches following the principles of hiding and masking, respectively. To hide the secret information in the power traces, the active fences try to consume power by operating the ROs appropriately, such that the overall power consumption is flat. To this end, the active fences are controlled by an on-chip RO-based power sensor. For instance, if the power sensor detects that the power consumption increases, the active fence decreases its power consumption to flatten the power changes. The other way to control the active fences is to use a pseudorandom number generator (PRNG). This approach follows the principle of masking, and it creates a noisy power profile for the attacker to measure. Experiments showed that by deploying active fences, the number of required power traces for a successful attack increases by 2 to 3 orders of magnitude. The authors also noted that the power sensor-activated fences using hiding principles are more effective in defenses.

\section{Securing Cloud Computation Using FPGAs}\label{sec:support}

On traditional cloud computation platforms, users can not control the underlying hardware that executes the computation. So, the users have to trust the cloud providers to handle their data properly and securely. Since the introduction of Intel SGX~\cite{costan2016intel}, the users do not have to trust the cloud provides. They need to trust only the manufacturers of the processors, i.e., Intel. This bottom-up trust model shows that one can minimize the trusted computing base to a trusted hardware like a processor. Cloud FPGAs, as programmable hardware in the clouds, can be used to construct an alternate trusted computing base for users when the processors are not trustworthy. We can use them as security monitors to check the behavior of the processor. Also, a programmable trusted peripheral allows one to prove certain security properties of a processor or an application. For example, a user can extract hardware fingerprints of cloud service instances and recover the architecture of the cloud infrastructures~\cite{tian2020fingerprinting}. Looking into the future, different architectures of FPGA clouds can emerge. For example, IBM Zurich is working on the \textit{cloudFPGA} project and proposing architectures for network-attached FPGAs~\cite{abel2017fpga}. This architecture allows the FPGA to access the network directly and process network packets independent of the processors.

\subsection{Architectural Supports}

In 2012, Eguro and others envisioned an FPGA-based cloud computing platform~\cite{eguro2012fpgas}. In this platform, the computation and data are sent to the cloud FPGA as encrypted data and an encrypted bitstream. In the cloud FPGA, the FPGA manufacturer pre-loads a secure bootstrapper bitstream. The secure bootstrapper shares a secret key with cloud clients and decrypts the data and bitstream files. After that, the data is computed using the hardware design described by the bitstream. Before the final results leave the cloud, they are encrypted again. This way, the client receives encrypted results, which he/she can decrypt locally.

To extend the idea of FPGA-based trusted computing in the cloud, Arasu \textit{et al.} proposed to use the above methodology for building an FPGA-based trusted co-processor, Cipherbase~\cite{arasu2013secure}. It aims at securing database operations. In the system architecture of Cipherbase, the whole database is encrypted, and the key is known to Cipherbase. Any expression evaluations are outsourced to Cipherbase. Because Cipherbase knows the secret keys, it can decrypt the data and evaluate the requested expression on the plaintext data easily. This architecture makes sure that the untrusted server which manages the database can have access only to the encrypted database, and the co-processor can assist the server in performing typical database operations efficiently. 

Xu \textit{et al.} ~\cite{xu2014pfc} proposed a similar idea as~\cite{eguro2012fpgas} to enable efficient privacy-preserving computation in the cloud. \cite{xu2014pfc} considered FPGAs as security containers and decrypts the data at the entry point of the cloud FPGA. Then, the data is encrypted again upon leaving the FPGAs. The main innovation in~\cite{xu2014pfc} is that the authors introduce proxy re-encryption to reduce the burden of key management in~\cite{eguro2012fpgas}. The concept of proxy re-encryption allows users to use their own keys for encryption. Moreover, the ciphertext encrypted by a user's key can be converted to a ciphertext of the same content encrypted by an FPGA's key without decrypting the ciphertext. This allows the user and the FPGA to exchange encrypted data without exchanging keys. 

\subsection{Cryptographic Accelerators}

The original purpose of introducing FPGAs into clouds was to enable the deployment of hardware accelerators to speed up the computation. Naturally, complex cryptographic operations need to be accelerated by hardware to satisfy the performance requirement in practice. In the rest of this subsection, we review two categories of cryptographic algorithms that demand to be implemented in hardware to improve their performance. The first one is fully/somewhat-homomorphic encryption~\cite{gentry2009fully,van2010fully,fan2012somewhat}, which was originally designed to allow computation over encrypted data in clouds. The second type of algorithms that needs hardware acceleration extensively is post-quantum cryptographic algorithms. These algorithms are the future direction of the cryptography for when quantum computers will be realized~\cite{chen2016report,bernstein2017post}. The third type is the privacy-preserving computation, such as garbled circuit evaluation~\cite{yao1986how}, which allows clouds to compute on user's program without leaking secrets. 

Many hardware accelerator architectures of homomorphic encryption have been proposed and implemented on FPGAs. For example,  ~\cite{poppelmann2015accelerating} proposed an architecture to accelerate the homomorphic encryption scheme YASHE. The hardware accelerator was able to provide roughly two orders of magnitude speedup compared with a software implementation of the same scheme by then. However, YASHE is no longer considered secure after an attack proposed in~\cite{albrecht2016subfield}. FV somewhat homomorphic encryption scheme~\cite{fan2012somewhat} was implemented in~\cite{roy2018hepcloud}, which requires 26.67 seconds for one homomorphic multiplication. Of the 26.67 seconds, only 3.36 seconds are used for actual computation, and the rest is the data access overhead due to the usage of a large parameter set. In fact, the large parameter size in secure homomorphic encryption schemes is the performance bottleneck on data transfers, as observed in~\cite{poppelmann2015accelerating,roy2018hepcloud,roy2019fpga}. With a smaller parameter setting, one recent work on accelerating the FV scheme shows 13 times performance improvement over a highly optimized software implementation of the FV scheme~\cite{roy2019fpga}. Also, researchers have addressed other performance bottlenecks as well. For instance, \cite{doroz2015accelerating,cilardo2016securing} proposed new multiplier architectures, and an improved Chinese Remainder Transformation accelerator was introduced in~\cite{cousins2016designing}.

Although the NIST standardization competition of post-quantum cryptography is ongoing, the hardware accelerators of the proposed signature schemes and key encapsulation schemes are being investigated in~\cite{wang2018fpga,basu2019nist}. \cite{wang2018fpga} presents a full FPGA implementation of a Niederreiter cryptosystem based on Binary Goppa Codes. This design achieves three orders of magnitude speedup compared to state-of-the-art software implementation. This shows a high potential of hardware accelerators for post-quantum cryptographic algorithms. More efforts have been presented in~\cite{basu2019nist}, which evaluate the potential performance gains of FPGA hardware implementations of NIST post-quantum competition candidates. The authors used HLS (High-level synthesis) tools for faster design-space exploration, and the results can be considered as a strong indicator of the hardware performance in the selection process of the NIST competition.

In addition, researchers have been implementing FPGA accelerators for privacy-preserving techniques, e.g., Garbled Circuits~\cite{yao1986how}, on the cloud. Garbled circuits, or secure function evaluation, is a technique that allows secure two-party computation. Using this, two mutually untrusted parties can jointly evaluate one function on their private inputs. MAXelerator was introduced as an FPGA accelerator on clouds for privacy-preserving machine learning in a garbled circuit form~\cite{hussain2018maxelerator}. The authors noticed that most of the privacy-sensitive computation in machine learning applications could be boiled down to multiply-accumulate operations, so they specifically used FPGAs to accelerate the multiply-accumulate garbled circuit computation. Overall, they achieved up to 57 times throughput improvement compared to the state-of-the-art software garbled circuit framework. A general-purpose FPGA accelerator for one party (garbler) in a two-party garbled circuit protocol was built by Huang \textit{et al.} in \cite{huang2019garbled}. They also implemented and demonstrated the FPGA accelerator on AWS instances, and they showed a 15 times speedup in garbler computation compared to a software-based garbled circuit framework.

\subsection{FPGA-based Security Primitives}

Having programmable hardware in clouds allows users to construct their security primitives in the clouds. For example, they can build their own physical unclonable functions (PUFs)~\cite{gassend2002silicon,herder2014physical} to identify individual FPGA chips. Many FPGA-based PUF designs were proposed and open-sourced online for fostering future research. As the first cryptographically secure PUF, LPN-based PUF~\cite{herder2016trapdoor} was implemented on a Zynq FPGA in a software/hardware co-design style, which perfectly fits the model of cloud FPGAs attached to a server~\cite{jin2017fpga}. The current state-of-the-art lightweight PUF design, interpose PUF (iPUF), which can resist all known attacks, was implemented in FPGAs as well~\cite{nguyen2019interpose}. One can extend PUFs to construct more security applications. For example, a PUF can be used for proof of execution in the cloud, which proves the identity of the device which runs the requested computation~\cite{gassend2008controlled}. Tian \textit{et al.} instantiated decay-based DRAM PUFs~\cite{xiong2016run} in AWS F1 instances, and thus they can fingerprint each instance in AWS cloud~\cite{tian2020fingerprinting}. Using these unique fingerprints of cloud instances, they experimentally figured out the probability of renting the same FPGA instance more than once, which provides unique insights for an attacker to launch further attacks.

Besides PUFs, true random number generators (TRNGs) can also be built on cloud FPGAs as an alternative to generating reliable randomness sources~\cite{petura2016survey}. Most of the FPGA-based TRNG designs take ROs as a core building block~\cite{fischer2008enhancing}, but, unfortunately, ROs, being combinational loops, are prohibited in the AWS clouds. One alternative approach is to exploit the metastability as a random source. For example, ~\cite{majzoobi2011fpga} presents a programmable delay line based TRNG design. This design does not require an RO as its building block. This demonstrates the possibility of building private and reliable TRNGs in the cloud.
\section{Related Recent Surveys}\label{sec:related} 

We acknowledge recent surveys on (cloud) FPGA security. \cite{zhang2019recent} surveyed attacks and defenses on FPGAs in general and did not focus on the unique characteristics of cloud FPGAs. \cite{mirzargar2019physical} focused on the security issues of cloud FPGAs, but the scope was narrowed down to only the side-channel attacks on cloud FPGAs, which is narrower than the scope of this paper. Similarly,~\cite{matasinvited} provided an excellent resource to study side-channel and fault injection attacks, but cloud FPGAs are facing more security issues than what they mentioned. \cite{trimberger2017security} in 2017 presented a high-level overview of the security issues in cloud FPGAs or FPGAs in data centers. However, as the security research on cloud FPGAs is evolving rapidly, remote side-channel attacks, fault-injections, and covert channels were not known by then, and therefore not covered in their paper. Nonetheless, some of the visionary countermeasures mentioned in~\cite{trimberger2017security} are shown to be effective in later research, e.g., the bitstream checkers and runtime monitoring. Last but not least, universities have started teaching courses on cloud FPGA~\cite{yale_course}. %
Such courses are an excellent starting point for researchers interested in cloud FPGAs.

Our paper not only presents recent work on cloud FPGA security in a comprehensive and timely manner but also surveys the related work on FPGAs to secure cloud computing. Thus, this survey covers the security issues beyond the attacks and defenses of FPGA security in clouds. 
\section{Conclusions and Future Challenges}\label{sec:future}

Cloud FPGA is an emerging trend with several security-related open problems awaiting exploration. From an attacker's point of view, we believe that the current known attacks in Section~\ref{sec:taxonomy} are still an incomplete list of possible attacks on cloud FPGAs. More attacks are likely to be proposed in the future, such as side-channel attacks exploiting other side-channel leakage sources. Also, using known information leakage sources, more information may be inferred.  Notice that researchers have implemented a power-based instruction disassembler on a microprocessor~\cite{park2018power}. This means that allowing attackers to observe the power consumption of the processor from the connected FPGA fabric can potentially reveal more information than what has been demonstrated in~\cite{zhao2018fpga,gravellier2019remote}.

In the system model, we demand strong security measures that can be used by cloud clients so that the clients do not need to trust the cloud providers. Ideally, the clients should be able to remotely verify some security properties (e.g., integrity) of their logic in the clouds. In addition, the clients may want to remotely verify the size of memory space using proof of space~\cite{ren2016proof}, the authenticity of the platform using remote attestation~\cite{costan2016intel}, the aliveness of the program using proof of aliveness~\cite{jin2019proof}, or the physical location of the data storage~\cite{bowers2011tell}. All of these traditional security issues in cloud computing need to be extended to include the FPGA platform as well. 

In a multi-tenant FPGA setting, more security mechanisms need to be in place to protect cloud users from other malicious users. How to efficiently defend against existing side-channel attacks and fault injection attacks without compromising the privacy of the user's design is still an open problem. Existing countermeasures fail to satisfy at least one of three requirements: low overhead, protection against known attacks, and the privacy of users. Traditional countermeasures provide strong security guarantees, but they incur significant hardware overhead. Bitstream checkers can detect malicious circuit designs given the list of known attacks. However, this requires access to cloud users' hardware designs. Passive online detection and active defenses require their deployment by the cloud providers and joint use to defend against side-channel and fault attacks. 

We can redesign the architecture of FPGA clouds in such a way that the FPGA can support the security features of the clouds and the FPGAs. For example, FPGAs can be used to support the secure boot of computing systems~\cite{pocklassery2018self}. Moreover, new FPGA cloud architectures can potentially limit the capability of malicious cloud providers. \cite{elnaggar2019multi,hategekimana2018secure,yazdanshenas2018improving} have effectively demonstrated new architectures by enforcing access control policy between processors and FPGA logic, or encrypting the messages transmitted between processors and FPGA logic. However, these mechanisms all require the cloud providers to deploy such a method by themselves, so it still requires some trust in the cloud providers. At least, trust in the infrastructure designers and manufacturers is needed if these mechanisms are built in the infrastructure hardware.

\begin{acks}
Chenglu Jin's research is supported in part by NYU CCS, NYU CUSP, and ONR grant N00014-18-1-2058. Ramesh Karri' research is supported in part by CCS-AD, NYU CCS, NSF awards 1526405 and 1513130, and ONR grant N00014-18-1-2058.
\end{acks}

\bibliographystyle{ACM-Reference-Format}
\bibliography{ref.bib}

\end{document}